# High-efficiency computational methodologies for electronic properties and structural characterization of Ge-Sb-Te based phase change materials

Shanzhong Xie, Kan-Hao Xue,* Shaojie Yuan, Shengxin Yang, Heng Yu, Rongchuan Gu, Ming Xu,* Xiangshui Miao

School of Integrated Circuits, Huazhong University of Science and Technology, Wuhan 430074, China

*Corresponding authors, email: xkh@hust.edu.cn (K.-H. X.), mxu@hust.edu.cn (M. X.).

## Abstract

Theoretical simulation to phase change materials such as Ge-Sb-Te has suffered from two methodology issues. On the one hand, there is a lack of efficient band gap correction method for density functional theory, which is suitable for these materials in both crystalline and amorphous phases, though the computational complexity should be kept at the local density approximation level. On the other hand, analysis of the coordination number in amorphous phases relies on an integration involving the radial distribution function, which adds to the complexity. In this work, we find that the shell DFT-1/2 method offers an overall band gap accuracy for phase change materials comparable to HSE06 hybrid functional, though its computational cost is around three orders of magnitude lower. Moreover, the mixed length-angle coordination number theory enables calculating the coordination numbers in the amorphous phase directly from the structure, with definite outcomes. The two methodologies could be helpful for high throuput simulation of phase change materials.

# I. Introduction

Chalcogenide-based phase change materials (PCMs), especially Ge-Sb-Te (GST) alloys, have garnered significant attention for applications in both optical storage and non-volatile electronic memory.[1–3] Alloys in the pseudo-binary line of GeTe-Sb$_2$Te$_3$ possess fast phase change speed, and various compositions exist, exemplified by Ge$_1$Sb$_4$Te$_7$ (GST-147), Ge$_1$Sb$_2$Te$_4$ (GST-124), Ge$_2$Sb$_2$Te$_5$ (GST-225) and Ge$_3$Sb$_2$Te$_6$ (GST-326). From GST-326 to GST-147, the switching speed gradually increases, but the corresponding stability of the amorphous phase degrades, and the crystallization temperature $T_x$ decreases to 85°C.[4] Theoretical simulation to PCMs mainly involves the dynamics of the phase change process, and the electronic structures. For the former, the microscopic mechanisms regarding the nucleation and growth processes are to be emphasized, where *ab initio* molecular dynamics (AIMD) is the core technique. The mechanisms that drive the phase transitions and the large resistivity differences between the crystalline and amorphous phase are still under intensive investigations. The latter task, studying the electronic structures from first principles, could serve as a foundation for the understanding of the physical properties of PCMs. The band gaps of both amorphous and (possibly) crystalline phases, as well as the trap states are the central topics.

Density functional theory[5,6] (DFT) has become the mainstream theoretical method to study the electronic structure of solids. While suffering from inaccurate exchange and correlation terms, it is much more efficient than post Hartree-Fock quantum chemistry methods, and is suitable for studying PCMs that typically require a large unit cell. Existing DFT studies have provided in-depth atomic scale understanding into the characteristics of GST compounds, especially GST-225. It is known that stable crystalline phases of GST are in either $P\bar{3}m1$ or $R\bar{3}m$ symmetries, where Ge, Sb, and Te atoms are stacked sequentially along the $c$-axis. Yet, the exact features of their electronic structures are still subject to certain controversies.[7–10] Accurate theoretical calculations on an entire series of GST-compounds should help to clarify these issues. Nevertheless, DFT suffers from a significant underestimation of band gaps under the normal local density approximation (LDA)[11,12] and generalized gradient approximation (GGA)[13–15] formulations. Two typical solutions are the hybrid functional scheme[16–18] and the quasi-particle approach within the GW approximation.[19–21]

While these methods effectively address the band gap issue, they inevitably introduce an increase of computational load by around 3 orders of magnitude for typical large-cell calculations. One still requires a band gap rectification method at the LDA-level of computational complexity.[22] The DFT-1/2 method as proposed by Ferreira, Marques and Teles in 2008,[23,24] and its variant, the shell DFT-1/2 method[25–28] as proposed in 2018, provide another route that only uses the self-energy potentials to correct the self-energy interaction errors due to LDA and GGA. In particular, shell DFT-1/2 could well recover the electronic band structure of Ge,[25] without referring to any empirical parameter, and it shows excellent band gap accuracy in Sb-based semiconductor superlattices for infrared-detection.[29] The computational speed of shell LDA-1/2 (which means LDA is used to account for the exchange-correlation part of DFT) is similar to conventional LDA,[30] and usually even faster than LDA[26] because the removal of unphysical electron self-interaction may accelerate the convergence of the self-consistent cycle. In shell DFT-1/2, a shell-like trimming function is used to confine the spatial range of the self-energy potential

$$\Theta(r) = \begin{cases} 0 & r < r_{in} \\ \left\{1 - \left[\frac{2(r - r_{in})}{r_{out} - r_{in}} - 1\right]^p\right\}^3 & r_{in} \leq r \leq r_{out} \\ 0 & r \geq r_{out} \end{cases}$$

where $p$ is an even integer of power index, which should be sufficiently large and is recommended to be $p = 20$. And $r_{in}$ and $r_{out}$ are the inner and outer radii of the cutoff function. Both values ought to be obtained from the variational principle, to maximize the band gap. This is because (shell) DFT-1/2 pulls down the valence band of the semiconductor or insulator, through rectifying the spurious electron self-interaction error. To recover the ground state from an ionized state, the total energy should be minimized, thus the band gap ought to be maximized. In this sense, the cutoff radii $r_{in}$ and $r_{out}$ should not be regarded as parameters in shell DFT-1/2, since there are calculated rather than from empirical data. After trimming, the self-energy potential is attached to the pseudopotentials of the anions for the self-energy correction for the valence band. In other words, the self-energy corrected pseudopotentials are used in standard self-consistent electronic structure calculations.[22,27] However, shell DFT-1/2 has not been applied to the calculations of GST yet.

A distinctive characteristic of PCMs is their inherent nature as narrow-bandgap semiconductors, which poses significant challenges for the precise determination and theoretical reproduction of their electronic structures. For instance, consider a computational method that could predict the band gaps of a certain type of semiconductors with an absolute error of 0.5 eV. While such predictive accuracy seems to be acceptable for a wide gap semiconductor (such as $Ga_2O_3$ with a ~4.9 eV gap) or an insulator (such as $HfO_2$ with a ~5.9 eV gap), this level of precision may prove inadequate for materials with narrower band gaps or more complex electronic structures (e.g., the band gaps of PCMs typically below 0.6 eV). Moreover, PCMs usually show a severe spin-orbit coupling (SOC) effect. For instance, GST contains a great proportion of heavy elements including Sb and Te. The works by Lawal *et al.*[31] and Hsieh et al.[32] show that, $Sb_2Te_3$ owns a certain time-reversal symmetry and strong SOC effect. It is reported to be a topological insulator with protected gapless surface states, rendering it different from other conventional semiconductors. In considering the SOC effect, the calculated band gaps ought to become much lower than non-SOC calculations.

It is sometimes tempting to believe that, the systematic band gap under-estimation by LDA/GGA could be compensated by the gap enhancement, through neglecting the SOC effect. In the literature, such calculations are very common and indeed some resulting band gaps seem to be reasonable. Lee and Jhi obtained a 0.26 eV band gap for crystalline GST-225 using GGA.[7] Shozo Yamanaka *et al.*[8] studied both the cubic and the hexagonal phases of GST-225 using LDA, and obtained a 0.1 eV band gap for the former, while the latter phase does not show a band gap in the calculation. Park *et al.*[9] calculated the electronic structures of GeTe, GST-147, GST-124, GST-225 as well as $Sb_2Te_3$, and the band gaps were, respectively, 0.66 eV, 0. 34eV, 0.43eV, 0.41 eV and 0.17 eV given by the Perdew-Burke-Ernzerhof (PBE) functional.[14] Ibarra-Hernández *et al.*[10] obtained 0.225 eV and 0.25 eV band gaps for GST-124 and GST-225, respectively, using the PBEsol (a solid version of PBE) functional.[33] Their calculation predicted GST-326 to be a metal with zero gap.

These results show that it is possible to obtain acceptable band gaps using conventional GGA while neglecting SOC, though in some cases it may fail to predict a finite band gap. Nevertheless, this scheme could be recommended only if the band gap under-estimation by GGA is exactly cancelled by the SOC-induced gap shrinkage. Unfortunately, they come from very distinct origins. The SOC effect is more severe for compounds with heavier elements, but the intrinsic gap under-estimation by GGA is not relevant to the nature of light or heavy elements. Hence, the compositional ratio of GeTe to $Sb_2Te_3$ ought to impact the amount of SOC-induced band gap reduction, thus conventional LDA/GGA alone is not supposed to predict reliable band gaps for the entire series of GST. In fact, even for the extensively studied GST-225, achieving a quantitatively accurate band gap is challenging when SOC is considered, particularly while the computational complexity is maintained at the LDA/GGA level.

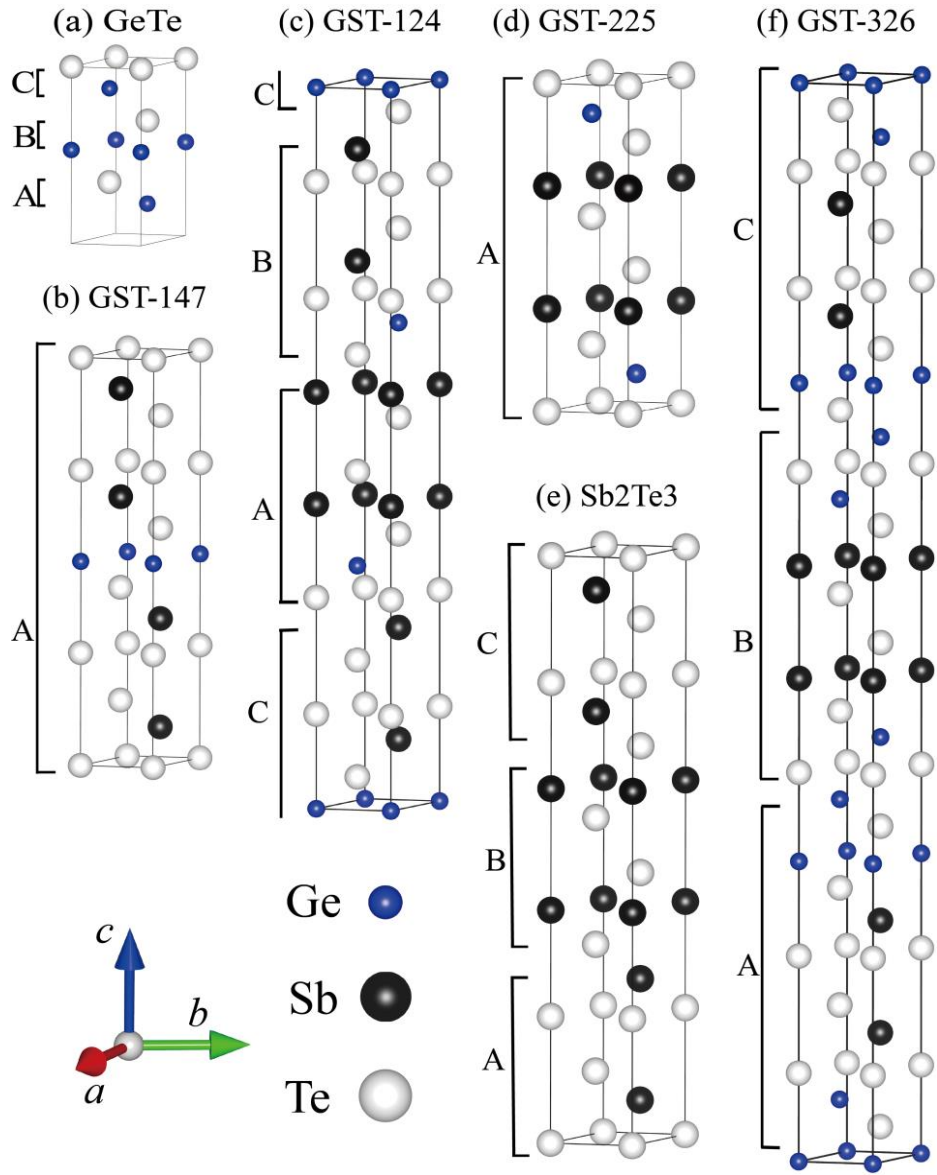

**Figure 1**. Illustration of the crystal structures for (a) GeTe; (b) GST-147; (c) GST-124; (d) GST-225; (e) $Sb_2Te_3$; and (f) GST-326. Each symbols A, B, or C denotes a basic repetitive stacking unit of the structure along c-axis.

This work, therefore, aims at exploring the computational method that is capable of recovering the band gaps of PCMs exemplified by GST, considering SOC but keeping the computational load at the same level of LDA/GGA. The focus is laid on the pseudo-binary line of GeTe-Sb$_2$Te$_3$ including GST-147, GST-124, GST-225, GST-326, as well as their parent materials Sb$_2$Te$_3$ and GeTe. The electronic structures of their crystalline and amorphous phases are both to be studied. A shell GGA-1/2 method with SOC considered (shGGA-1/2+SOC for short) is shown to be highly efficient, with comparison to GGA, GGA+SOC, shGGA-1/2, HSE06,[18,34,35] as well as HSE06+SOC. Moreover, for coordination number analysis, we have found that the mixed length-angle coordination number theory (MLAC)[36] is particularly suitable for PCMs.

**Table 1.** Structural information and self-energy correction schemes for the six materials under investigation

| Material | Space group | Lattice constant (Å) | | | | Exact way of shGGA-1/2 |
|---|---|---|---|---|---|---|
| | | $a_0$ | | $c_0$ | | |
| Sb$_2$Te$_3$ | $R\bar{3}m$ (No. 166) | 4.34[a] | 4.26[b], 4.27[c], 4.34[i] | 31.44[a] | 30.45[b], 30.45[c], 31.29i | shGGA-1/4-1/4 |
| GST-147 | $P\bar{3}m1$ (No. 164) | 4.23[a] | 4.24[j] | 24.05[a] | 23.76j | shGGA-0-0-1/2 |
| GST-124 | $R\bar{3}m$ (No. 166) | 4.23[a] | 4.27[d], 4.25[h], 4.25[h] | 41.26[a] | 41.7[d], 41.0[h], 41.0h | shGGA-0-0-1/2 |
| GST-225 | $P\bar{3}m1$ (No. 164) | 4.21[a] | 4.22[e], 4.25[h] | 17.16[a] | 17.24[e], 18.27[h] | shGGA-0-0-1/2 |
| GST-326 | $R\bar{3}m$ (No. 166) | 4.21[a] | 4.21[f], 4.25[h] | 61.76[a] | 62.31[f], 62.6[h] | shGGA-0-0-1/2 |
| GeTe | $R3m$ (No. 160) | 4.23[a] | 4.17[g], 4.23[k] | 10.86[a] | 10.62[g], 10.92[k] | shGGA-0-1/2 |

[a] The calculation result of our own.
[b] Experiment in Ref. [37].
[c] Experiment in Ref.[38].
[d] Experiment at 873 K in Ref.[39].
[e] Experiment in Ref.[40].
[f] Experiment at 90 K in Ref.[41].
[g] Experiment in Ref.[42].
[h] Experiment in Ref.[43].
[i] Experiment in Ref.[44].
[j] Experiment in Ref.[45].

## II. Structural models and computational settings

**Table 1** and **Figure 1** demonstrate the structural information of the six materials. Crystalline GST-225 has a hexagonal symmetry with space group $P\bar{3}m1$, whose basic repetitive stacking unit is Te-Ge-Te-Sb-Te-Te-Sb-Te-Ge, including 9 layers. The reference experimental lattice constants are a = 4.25 Å, c = 18.27 Å,[43] which are used for setting up the model cell. GST-124 is short of a Ge-Te bi-layer in its basic stacking unit compared with GST-225, *i.e.*, Te-Ge-Te-Sb-Te-Te-Sb. In order to maintain the periodicity and structural stability, the supercell for calculation has to be dividable by 3 along the *c*-axis. Hence, we set up a $1 \times 1 \times 3$ supercell that contains 21 layers, with initial lattice parameters as $a$ = 4.25 Å, $c$ = 41.00 Å.[43] The basic stacking unit of GST-326, on the other hand, contains Te-Ge-Te-Sb-Te-Te-Sb-Te-Ge-Te-Ge. It owns add additional bilayer of Ge-Te compared with GST-225. Hence, a 33-layer model supercell had to set up for GST-326, initially with $a$ = 4.25 Å, c = 62.60 Å.[43] The basic stacking unit for GST-147 consists of Te-Sb-Te-Te-Sb-Te-Ge-Te-Sb-Te-Te-Sb, with 12 layers and $a$ = 4.236 Å, $c$ = 23.761 Å.[45] Matsunaga et al.[45] revealed a van der Waals-like weak force between Te and Te layers, through the X-ray diffraction method as well as DFT calculations. This necessitates the van der Waals force correction in our calculations. Sb$_2$Te$_3$ has a hexagonal lattice with space group $R\bar{3}m$, whose basic stacking unit is Te-Sb-Te-Te-Sb. To let 3 divides the number of layers along *c*-axis, a 15-layer supercell was thus established with $a$ = 4.34 Å, $c$ = 31.29 Å.[46] At room temperature, GeTe also shows a hexagonal lattice with the $R3m$ space group. Its basic stacking unit contains merely Te-Ge, but the supercell has to contain 6 layers with $a$ = 4.23 Å, $c$ = 10.92 Å.[46]

DFT calculations were carried out using the Vienna Ab initio Simulation Package (VASP 5.4.4),[47,48] using the projector augmented-wave[49,50] (PAW) method with a 350 eV plane-wave kinetic energy cutoff. The exchange-correlation energy was treated within the generalized gradient approximation using the Perdew-Burke-Ernzerhof functional.[14] The valence electron configurations were: 4s and 4p for Ge; 5s and 5p for Sb and Te; 3s and 3p for Al; 3d, 4s and 4p for Ga; 4d, 5s and 5p for In. The van der Waals force correction was carried out using the DFT-D2 scheme by Grimme.[51] Structural optimization criteria were, (i) the residual stress in any direction was less than 100 MPa; (ii) the Hellmann-Feynman force for any atom was below 0.005 eV/Å in any direction. Equal-spacing

Monkhorst-Pack[52] k point meshes were used to sample the Brillouin zones, with detailed k point information is given in **Table S1**.

For electronic structure calculations, the shell DFT-1/2 method attaches the self-energy potentials to the anions, which account for a majority part of the valence band states. For highly ionic compounds it is straightforward to identify the anion elements. Yet, the bonding in PCMs is different from that of a typical ionic bond. Hence, we adopted the differential charge method to explore the spatial location of valence band holes. This involves subtracting 0.01 electron from a unit cell, and compare the charge distribution between the neutral cell and that of the ionized cell. The reason for using 0.01 electron instead of one electron lies in that this does not perturb the electronic states to an undesirable extent. The differential charge density will then be magnified by 100 times to recover one electron removal.[25] The hole locations are illustrated in **Figure S1**. It seems that for GeTe one should prefer shGGA-0-1/2, where 0 and 1/2 are the amounts of equivalent electron removal from Ge and Te, respectively. For $Sb_2Te_3$, on the other hand, shGGA-1/4-1/4 is to be carried out, where Sb and Te are both subject to 1/4 electron removal. The various GST models (GST-147, GST-124, GST-225, GST-326) fit shGGA-0-0-1/2, where 0, 0 and 1/2 are the amounts of equivalent electron removal from Ge, Sb and Te, respectively. Through scanning $r_{in}$ and $r_{out}$ to maximize the band gaps, all cutoff radii are obtained unambiguously as listed in **Table 2**. Since near the extreme point, the band gap varies very slowly with respect to the cutoff radii, it turns out that a consistent setting Te $r_{in} = 0.9$ Bohr and Te $r_{out} = 3.0$ Bohr can used for GST in general, though in GST-326 we used the optimal value $r_{out} = 2.9$ Bohr, which makes extremely little difference.

A melt-quench scheme[53] was employed to generate the amorphous structures of $Sb_2Te_3$ (a-$Sb_2Te_3$), amorphous GST (a-GST-147, a-GST-124, a-GST-225 as well as a-GST-326), and amorphous GeTe (a-GeTe). The AIMD simulation was based on the second-generation Car–Parrinello scheme[54] in the canonical ensemble (NVT) with a stochastic Langevin thermostat.[55] The time step was set to 2 fs, and only the Γ point was used to sample the Brillouin zone of all models. All model supercells were first melted at a high temperature of 2000 K for 20 ps. Subsequently, during the quenching process, the size of the simulation box was adjusted multiple times, to minimize internal stress as much as possible. The models were eventually quenched to 300 K, and the resulting amorphous structures underwent thorough geometric relaxation to further reduce internal stress, ensuring that the absolute value of the final stress in all models was less than 100 MPa.

**Table 2**. The optimized self-energy cutoff radii for the six materials under shell GGA-1/2 calculations.

|  |  | Radius (Bohr) | | | | | |
|---|---|---|---|---|---|---|---|
|  |  | $Sb_2Te_3$ | GST-147 | GST-124 | GST-225 | GST-326 | GeTe |
| $r_{in}$ | | Sb: 0.1 | Te: 0.8 | Te: 0.9 | Te: 0.9 | Te: 0.9 | Te: 0.8 |
|  | | Te: 1.2 | | | | | |
| $r_{out}$ | | Sb: 1.6 | Te: 3.0 | Te: 3.0 | Te: 3.0 | Te: 2.9 | Te: 3.0 |
|  | | Te: 3.4 | | | | | |

## III. Electronic structure and bonding in the crystalline phases

The electronic band structures for the crystalline models are illustrated in **Figures 2-4** as well as in **Table 3**, using GGA, GGA+SOC, shGGA-1/2, shGGA-1/2+SOC, HSE06, and HSE06+SOC respectively. Using conventional GGA (**Figures 2(a)-2(f)**), $Sb_2Te_3$ is shown to possess a direct 0.18 eV band gap, with the valence band maximum (VBM) and conduction band minimum (CBM) both lying at Γ. GST-147 has a 0.24 eV direct gap, both its VBM and CBM reside at A. Near Γ, however, GST-147 has two valleys of the conduction band, whose energies are almost degenerate with the Γ point (differing by merely 0.2 meV). For GST-124 and GST-225, GGA predicts indirect gaps. Their VBMs both lie at Γ, and the CBMs are close to Γ. The magnitudes of GGA gaps are 0.35 eV and 0.24 eV, respectively. GST-326 is predicted to possess a direct Γ—Γ gap of 0.29 eV. In GeTe, the VBM is along the Γ—L line, while the CBM is very close to the T point. The indirect gap value is 0.54 eV as predicted by plain GGA.

When SOC is turned on, it is observed from **Figures 2(g)-2(l)** that all band gaps become less than 0.1 eV, except for GeTe. And in GST-147, the type of band gap has changed from direct to indirect. However, the GST-124 and GST-225 now show direct gaps after SOC is considered, though they were predicted as direct semiconductors by GGA. Note that these materials possess narrow gaps, and in each case the difference between direct and indirect gaps is not severe. Hence, it is difficult to precisely judge the type of band gaps from experimental, and there is hardly any measurement of the type of gap published. Hence, the magnitude of band gap should be our primary focus here, though there is clue from our theoretical calculation that turning on SOC could change the type of gap in GST.

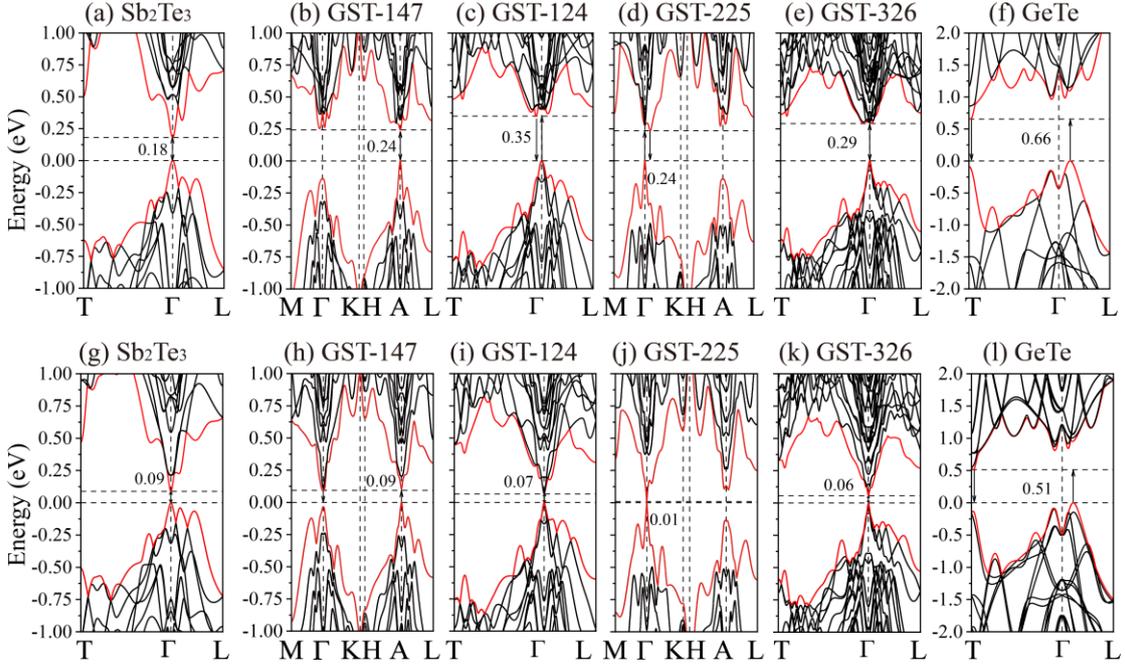

**Figure 2.** Energy band diagrams calculated using GGA. (a) $Sb_2Te_3$ without SOC; (b) GST-147 without SOC; (c) GST-124 without SOC; (d) GST-225 without SOC; (e) GST-326 without SOC; (f) GeTe without SOC; (g) $Sb_2Te_3$ with SOC; (h) GST-147 with SOC; (i) GST-124 with SOC; (j) GST-225 with SOC; (k) GST-326 with SOC; (l) GeTe with SOC.

Both the self-energy correction method shGGA-1/2 and the hybrid functional HSE06 are very effective in rectifying the band gap problem of GGA. **Figure 3** and **Figure 4** show the shGGA-1/2 and HSE06 results, respectively, either without or with SOC turned on. In the absence of SOC effect, the predicted gap values in (shGGA-1/2, HSE06) format are $Sb_2Te_3$ (0.67 eV, 0.87 eV), GST-147 (0.89 eV, 0.87 eV), GST-124 (0.95 eV, 0.98 eV), GST-225 (0.89 eV, 0.83 eV), GST-32 (0.90 eV, 0.87 eV) and GeTe (1.43 eV, 1.39 eV). The shGGA-1/2 values are all close to HSE06 values, even though shGGA-1/2 is computationally much lighter. Considering the SOC effect, however, the band gaps are predicted to be $Sb_2Te_3$ (shGGA-1/2: 0.27 eV, HSE06: 0.51 eV), GST-147 (0.45 eV, 0.42

eV), GST-124 (0.59 eV, 0.58 eV), GST-225 (0.57 eV, 0.54 eV), GST-326 (0.61 eV, 0.63 eV), GeTe (1.31 eV, 1.21 eV). The only big discrepancy between shGGA-1/2 and HSE06 occurs in the case of $Sb_2Te_3$, but the shGGA-1/2 gap is closer to experimental. The SOC effect also has a great impact in shGGA-1/2 and HSE06 calculations. For $Sb_2Te_3$ and GeTe, shGGA-1/2+SOC predicts a 0.27 eV direct gap and a 1.31 eV indirect gap, respectively. The result for $Sb_2Te_3$ is consistent with a $GW$ calculation by Lawal et al.[56] (0.22 eV). Without considering SOC, the shGGA-1/2 gap of $Sb_2Te_3$ is 0.67 eV, thus the SOC-induced gap shrinkage is as large as 0.4 eV. The strong SOC effect observed in $Sb_2Te_3$ agrees with the experimental results.[31,32]

With more GeTe contained in GST, it is discovered that the shGGA-1/2+SOC gaps show an increasing trend, which is consistent with the experimental results of Park et al.[9] A comparison of shGGA-1/2+SOC gaps with typical experimental gaps is in general satisfactory: GST-225 (shGGA-1/2+SOC: 0.57 eV, experimental: 0.57 eV); GST-147 (shGGA-1/2+SOC: 0.59 eV, experimental: 0.55 eV). A benchmark for GST-326 is not yet possible due to a lack of experimental data. In addition, since Te is the heavier anion element in GST, it is supposed that the effect of SOC will be more severe for GST with more Te content. As shown in **Table 4**, both shGGA-1/2 and HSE06 calculations follow this trend exactly. This is verified by computing $E_g - E_g^{SOC}$, where a larger difference indicates a stronger SOC effect. Nevertheless, plain GGA calculations do not follow this trend, possibly due to the over-narrow gap nature in this series of compounds. This also confirms the point that, GGA-induced gap under-estimation cannot simply compensate the SOC-induced gap variation in GST, because the two mechanisms intrinsically have very different origins. Provided that the HSE06+SOC results are regarded as the reliable standards, the through inspecting $E_g^{GGA+SOC} - E_g^{HSE06+SOC}$, one could find that (i) GGA+SOC leads to very inaccurate band gap values; (ii) GGA+SOC does not show a consistent trend across the GST compositions, since GST-147 behaves quite differently compared with other compounds. On the contrary, $E_g^{shGGA-\frac{1}{2}+SOC} - E_g^{HSE06+SOC}$ shows a very consistent trend as that of HSE06 (also shown in **Table 4**).

While similar in electronic structure accuracy, the computational efficiencies of shGGA-1/2 and HSE06 differ much. To better quantify such difference, we tested the computational time of all six computational methods (GGA, GGA+SOC, shGGA-1/2, shGGA-1/2+SOC, HSE06, HSE06+SOC) in the selected series of PCM materials. Usually, less $k$ points have to be used in HSE06 calculations, but GGA and shGGA-1/2 permit more $k$ points where the time cost is still tolerable. This inevitably causes an unfair comparison if only the total time is recorded. Hence, we re-did the GGA and shGGA-1/2 type calculations using the same $k$ point settings as in the corresponding hybrid functional calculations. However, with SOC the number of k points is still too many for HSE06, thus in several HSE06+SOC calculations we used less $k$ points. Another issue lies in that each material could be subject to a specific number of irreducible $k$ points. Hence, we emphasize the average time cost per $k$ point. All the calculations were carried out using a 72-core computer, and the data are listed in **Table 5**. It turns out clearly that shGGA-1/2 owns the same computational speed as conventional GGA, regardless of whether SOC is turned on or not. On the other hand, an HSE06 calculation typically has a time cost of nearly three orders of magnitude higher. Detailed $k$ point information during the tests can be found in **Supplementary Note 1**.

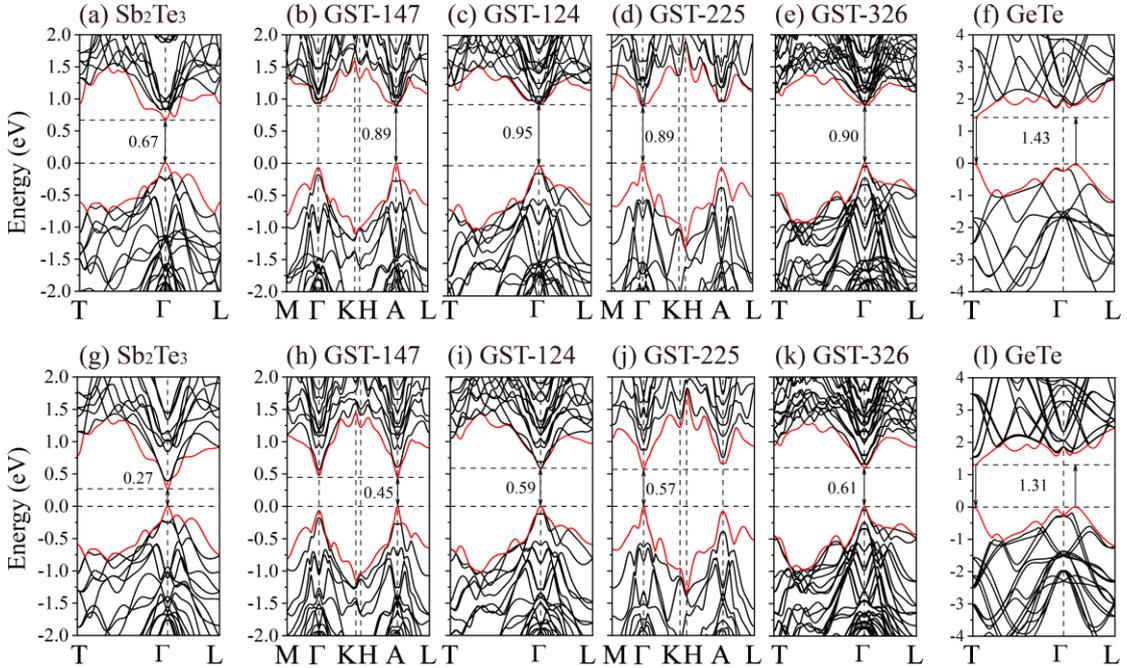

**Figure 3.** Energy band diagrams calculated using shGGA-1/2. (a) $Sb_2Te_3$ without SOC; (b) GST-147 without SOC; (c) GST-124 without SOC; (d) GST-225 without SOC; (e) GST-326 without SOC; (f) GeTe without SOC; (g) $Sb_2Te_3$ with SOC; (h) GST-147 with SOC; (i) GST-124 with SOC; (j) GST-225 with SOC; (k) GST-326 with SOC; (l) GeTe with SOC.

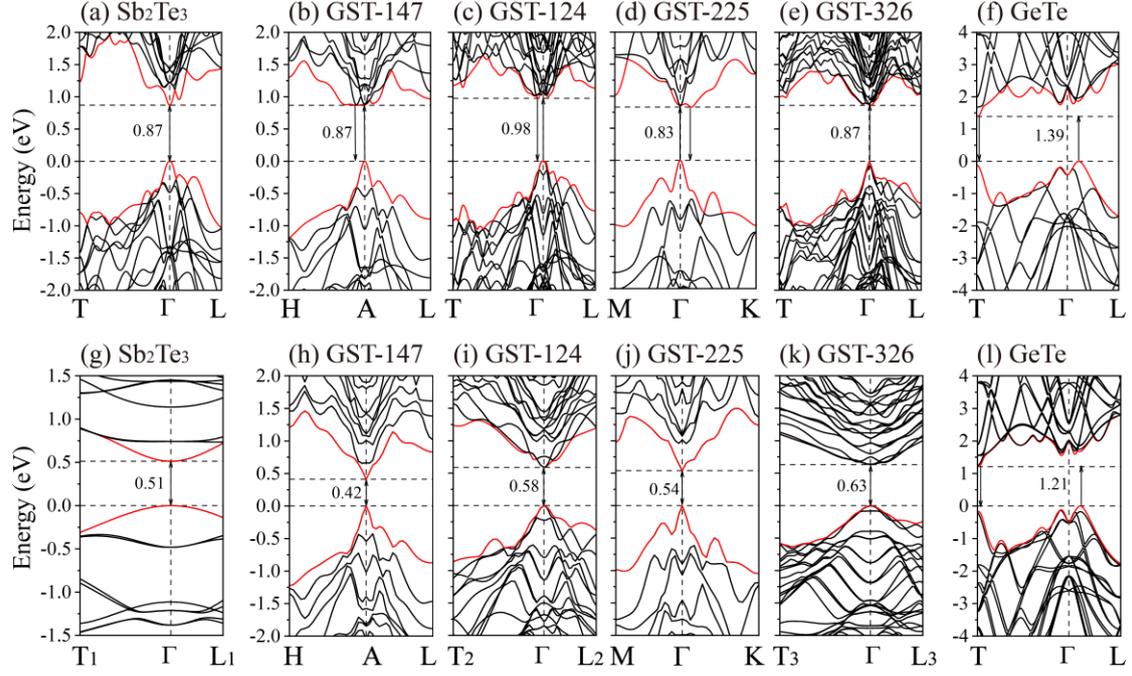

**Figure 4.** Energy band diagrams calculated using the HSE06 hybrid functional. (a) $Sb_2Te_3$ without SOC; (b) GST-147 without SOC; (c) GST-124 without SOC; (d) GST-225 without SOC; (e) GST-326 without SOC; (f) GeTe without SOC; (g) $Sb_2Te_3$ with SOC; (h) GST-147 with SOC; (i) GST-124 with SOC; (j) GST-225 with SOC; (k) GST-326 with SOC; (l) GeTe with SOC. Here $T_1$, $T_2$ and $T_3$ are three ordinary $k$ points along the Γ-T line; $L_1$, $L_2$ and $L_3$ are three ordinary $k$ points along the Γ-L line.

**Table 3.** Calculated and experimental band gaps of $Sb_2Te_3$, GST-147, GST-124, GST-225, GST-326 and GeTe, where *d* and *i* indicate direct and indirect gaps, respectively.

|  | Band gap (eV) | | | | | |
|---|---|---|---|---|---|---|
|  | $Sb_2Te_3$ | GST-147 | GST-124 | GST-225 | GST-326 | GeTe |
| $E_g^{GGA}$ | 0.18 (*d*) | 0.24 (*d*) | 0.35 (*d*) | 0.24 (*i*) | 0.29 (*i*) | 0.66 (*i*) |
| $E_g^{GGA+SOC}$ | 0.09 (*d*) | 0.09 (*i*) | 0.07 (*d*) | 0.01 (*d*) | 0.06 (*d*) | 0.51 (*i*) |
| $E_g^{shGGA-\frac{1}{2}}$ | 0.67 (*d*) | 0.89 (*d*) | 0.95 (*d*) | 0.89 (*d*) | 0.90 (*d*) | 1.43 (*i*) |
| $E_g^{shGGA-\frac{1}{2}+SOC}$ | 0.27 (*d*) | 0.45 (*d*) | 0.59 (*d*) | 0.57 (*d*) | 0.61 (*d*) | 1.31 (*i*) |
| $E_g^{HSE06}$ | 0.87 (*d*) | 0.87 (*i*) | 0.98 (*i*) | 0.83 (*i*) | 0.87 (*d*) | 1.39 (*i*) |
| $E_g^{HSE06+SOC}$ | 0.51 (*d*) | 0.42 (*d*) | 0.58 (*d*) | 0.54 (*d*) | 0.63 (*d*) | 1.21 (*i*) |
| Experimental | 0.15—0.22[57] | - | 0.55[9] | 0.57,[9] 0.5[58] | - | 0.61[9] |

**Table 4.** Impact of SOC on the band gaps of the crystalline GST samples

|  | GST-326 | GST-225 | GST-124 | GST-147 |
|---|---|---|---|---|
| Te content | 54.5% | 55.6% | 57.1% | 58.3% |
| $E_g^{GGA} - E_g^{GGA+SOC}$ (eV) | **0.23** | **0.23** | **0.28** | **0.15** |
| $E_g^{shGGA-\frac{1}{2}} - E_g^{shGGA-\frac{1}{2}+SOC}$ (eV) | **0.29** | **0.32** | **0.36** | **0.44** |
| $E_g^{HSE06} - E_g^{HSE06+SOC}$ (eV) | **0.24** | **0.29** | **0.40** | **0.45** |
| $E_g^{GGA+SOC} - E_g^{HSE06+SOC}$ (eV) | -0.57 | -0.53 | -0.51 | -0.33 |
| $E_g^{shGGA-\frac{1}{2}+SOC} - E_g^{HSE06+SOC}$ (eV) | -0.02 | 0.03 | 0.01 | 0.03 |

In characterizing the local structure in GST, the distribution of Sb-Te bong lengths usually shows two peaks at around 2.97 Å and 3.17 Å.[59–61] Coordination number (CN) for an atom/ion in a solid has traditionally been determined according to the lengths of its bonds. This method has certain limitations in case the bond length distribution is not sharply divided into short and long classes. Although the 0.2 Å difference in the bond length of crystalline GST does not cause confusion, identifying the CN in amorphous GST cannot simply rely on the bond length analysis, but it requires an integration taking advantage of the pair correlation function. Recently, we have proposed a CN theory based on both the bond lengths and bond angles, which may be called the mixed length-angle coordination (MLAC). This theory also lists the bond lengths in an ascending order, but whether a new bond is counted in the coordination depends on its angles to existing bonds. If it makes an angle greater than $\theta=65°$ with respect to any of the existing bonds, then it is counted. Otherwise, the counting stops. Detailed explanation of the MLAC theory is given in the origin publication, as well as in **Supplementary Note 2** of this work. Using the MLAC method, we find that the CNs of Ge or Sb are always 6 in the six materials under investigation, but the CNs of Te are differing. According to the data shown in **Table S4**, three distinct Te sites can be identified. Te1 represents a Te atom with 6-coordination (6C), which only has bonds with Ge, or only with Sb. Te2 is a different Te site that is 6C, but it possesses Ge-Te bonds and Sb-Te bonds, simultaneously. Te3 is a 3C Te site with connection to Sb atoms only. Detailed information regarding the coordination configurations in the six materials can be found in **Supplementary Note 3**.

**Table 5.** Average time costs for one irreducible k-point in each method

|  | Time (s) | | | | | |
| --- | --- | --- | --- | --- | --- | --- |
|  | GGA | GGA+SOC | shGGA-1/2 | shGGA-1/2+SOC | HSE06 | HSE06+SOC |
| $Sb_2Te_3$ | 1.14 | 2.90 | 1.10 | 3.21 | 8061.34 (2.24 h) | 20528.86 (5.70 h) |
| GST-147 | 2.21 | 1.83 | 2.00 | 1.50 | 12225.77 (3.40 h) | 25301.10 (7.03 h) |
| GST-124 | 2.47 | 10.62 | 2.46 | 10.54 | 15159.43 (4.21 h) | 11622.23 (3.23 h) |
| GST-225 | 0.34 | 0.67 | 0.29 | 0.81 | 5878.82 (1.63 h) | 5044.05 (1.40 h) |
| GST-326 | 3.07 | 29.90 | 3.00 | 33.18 | 22104.04 (6.14 h) | 69231.04 (19.23 h) |
| GeTe | 0.08 | 0.27 | 0.08 | 0.31 | 1087.10 (0.30 h) | 1169.47 (0.32 h) |

As demonstrated in **Table S7**, there are two sorts of Sb-Te bonds in $Sb_2Te_3$ and the various GST models. In GST the bond lengths are 2.99 Å and 3.15 Å, respectively. In $Sb_2Te_3$ the bonds are slightly longer, but with the same trend discovered. In GeTe, there are two distinct Ge-Te bond lengths as well. This fact is consistent with many reports in the literature. For example, based on *ab initio* Raman spectra, Sosso *et al.* revealed that $Sb_2Te_3$ involves two Sb-Te bond length values of 2.97 Å and 3.17 Å.[59] Kolobov *et al.* revealed from extended X-ray absorption fine structure spectroscopy that there are two sorts of Sb-Te bong lengths in GST-225, with values of 2.83 Å and 3.15 Å, respectively.[60] And it was reported experimentally that GeTe involves two bond lengths, 2.80 Å and 3.13 Å.[60,61] In addition, for a 6C Sb atom in GST or $Sb_2Te_3$, its six bonds can be divided into 3 short bonds and 3 long bonds. A short bond involves bonding to a 3C Te atom (*i.e.*, Te3), while a long bond involves bonding to a 6C Te atom (*i.e.*, Te1 or Te2).

## IV. Analyses to the amorphous phases

The amorphous models for the six materials are demonstrated in **Figure 5** as well as **Table S8**. The lattice parameters and theoretical number densities are taken from the fully relaxed structures, though the initial structures are meta-stable cubic GST. In a such initial structure, Te atoms constitute a sub-lattice of the rock salt structure, while Ge, Sb and vacancies are randomly distributed on the other sub-lattice.[40,42,62] In each of the six materials under investigation, the percentage of vacancies is controlled as ~20%. The total atoms in a supercell is between 270 and 300. The theoretical atomic number densities range between 0.0272 Å$^{-3}$ and 0.0310 atoms Å$^{-3}$ in these models, close to experimental value 0.030 atoms Å$^{-3}$.[63] At 300 K, the partial pair distribution functions $g(r)$ are shown in **Figure S9**. The definition of the partial pair distribution function $g(r)$ and its calculation method are explained in **Supplementary Note 4**. With more GeTe content in GST, the $g(r)$ peak for the Ge-Te bond obviously grows, verifying the proper structures of the amorphous models. According to **Figure 5**, no substantial de-mixing is observed, and all our generated amorphous models still represent homogeneous phases.

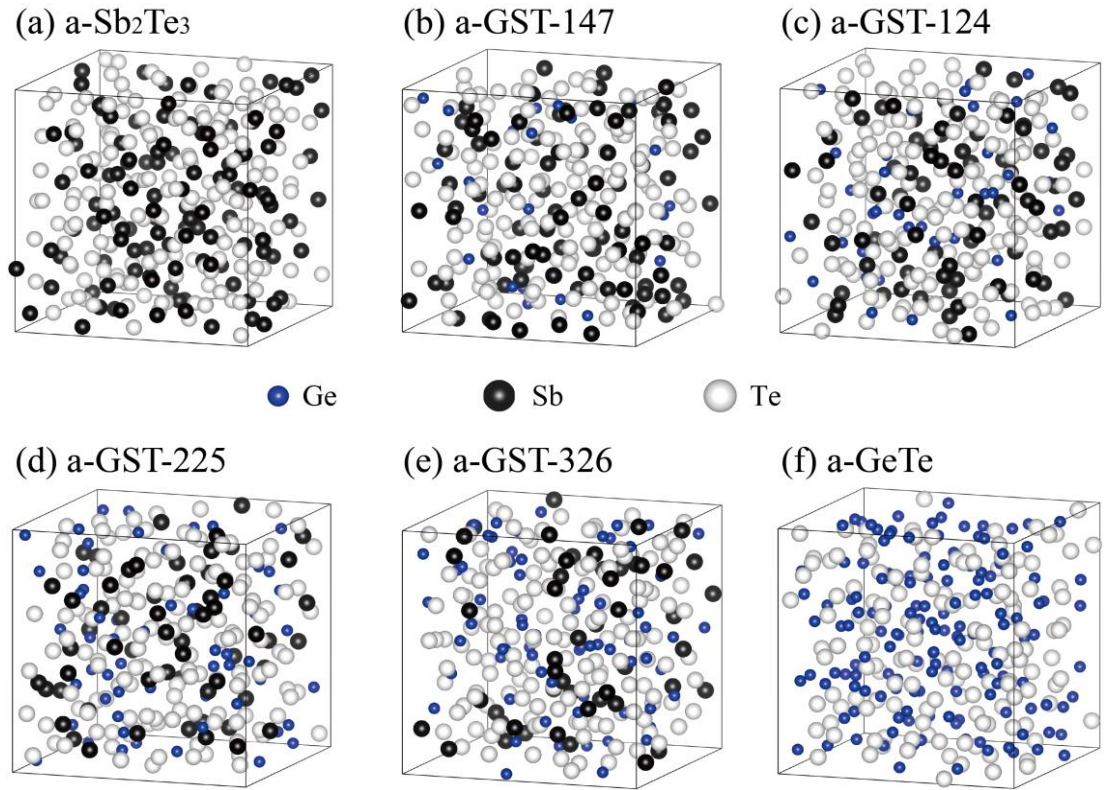

**Figure 5.** Model structures of (a) a-Sb$_2$Te$_3$; (b) a-GST-147; (c) a-GST-124; (d) a-GST-225; (e) a-GST-326; and (f) a-GeTe.

We then calculated the electronic structures of these amorphous models, using GGA, GGA+SOC, shGGA-1/2, as well as shGGA-1/2+SOC. Hybrid functionals were not applied due to the high computational cost. We used the inverse participation ratio (IPR) to identify the mobility gaps and trap states. The definition of the mobility gap and the calculation method for IPR are given in **Supplementary Note 5** and **Supplementary Note 6**. In general, larger IPR values indicate more strongly localized electron states. The mobility gaps ($E_{gm}$) of the amorphous models were obtained through calculating the energy separation between the mobility edges, defined by relatively lower IPR values of valence-band and conduction-band states compared with the trap states in the forbidden band.[64] The most common composition GST-225 is taken as our focus for analysis. As illustrated in **Figure 6**, the roughly estimated $E_{gm}$ values by GGA, GGA+SOC, shGGA-1/2 and shGGA-1/2+SOC are 0.69 eV, 0.61 eV, 0.89 eV and 0.81 eV, respectively. All methods predict some trap states. These trap states show large IPR values, indicating that the carriers trapped at these localized states will contribute little to the electrical conduction at room temperature because of the low mobility. It turns out that shGGA-1/2 could capture more gap states compared with GGA. Experimentally, Kato et al.[65] obtained a 0.74 eV Tauc gap for a-GST-225, and the energy width of the Urbach edge was ~0.2 eV (relative to the Fermi level, which was set to zero). They further pointed out that the exact location of the low-energy edge was unknown, and accordingly, the density of states around the valence band maximum (VBM) was vague. For sulfide and selenide glasses, it is known that the Tauc gap is smaller than the mobility gap and that the Urbach edge is governed by the valence-band tail.[66] Hence, the VBM should be lower than the Fermi level by at least 0.2 eV, according to experimental clues. And the mobility gap should be larger than 0.74 eV. In our calculation results, the distance of the Fermi level with respect to the lower edge of the mobility gap is indicated as A, B, C and D in each case. The distance values are 0.16 eV (GGA), 0.20 eV (GGA+SOC), 0.16 eV (shGGA-1/2) and 0.26 eV (shGGA-1/2+SOC), respectively. Of them, only that predicted by shGGA-1/2+SOC is greater than 0.2 eV. The quality of shGGA-1/2+SOC calculation has been demonstrated in terms of the band edge location, trap state and the localization effect.

On account of the effectiveness of shGGA-1/2+SOC in recovering the electronic structures of a-GST-225, we finished the calculations for other amorphous model structures. As shown in **Figure 7**, the mobility gaps predicted by shGGA-1/2+SOC for a-$Sb_2Te_3$, a-GST-147, a-GST-124, a-GST-326, a-GeTe are 0.67 eV, 0.76 eV, 0.79 eV, 0.83 eV and 0.88 eV, respectively, showing a consistent trend that is consistent with the experimental mobility gap results by Park *et al.*[9]

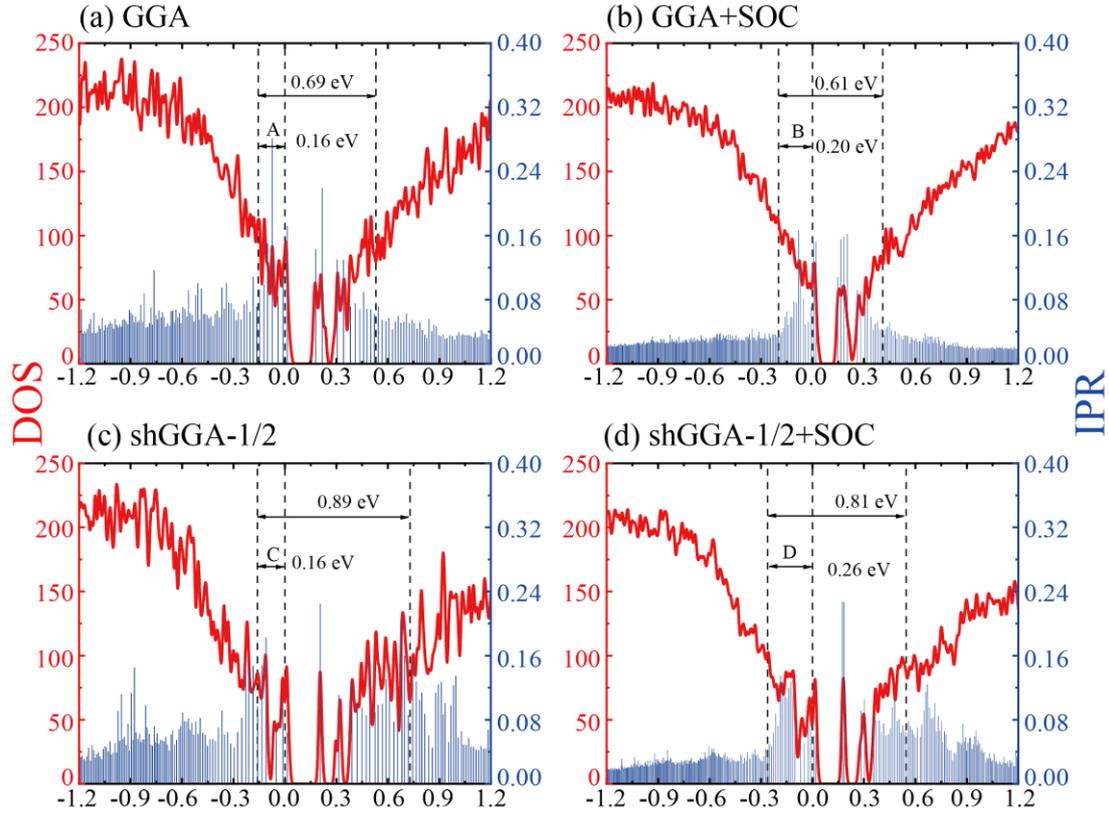

**Figure 6.** The density of states as well as the corresponding normalized IPR for a-GST-225, calculated using various methods. (a) GGA; (b) GGA+SOC; (c) shGGA-1/2; (d) shGGA-1/2+SOC. The Fermi level corresponds to zero energy.

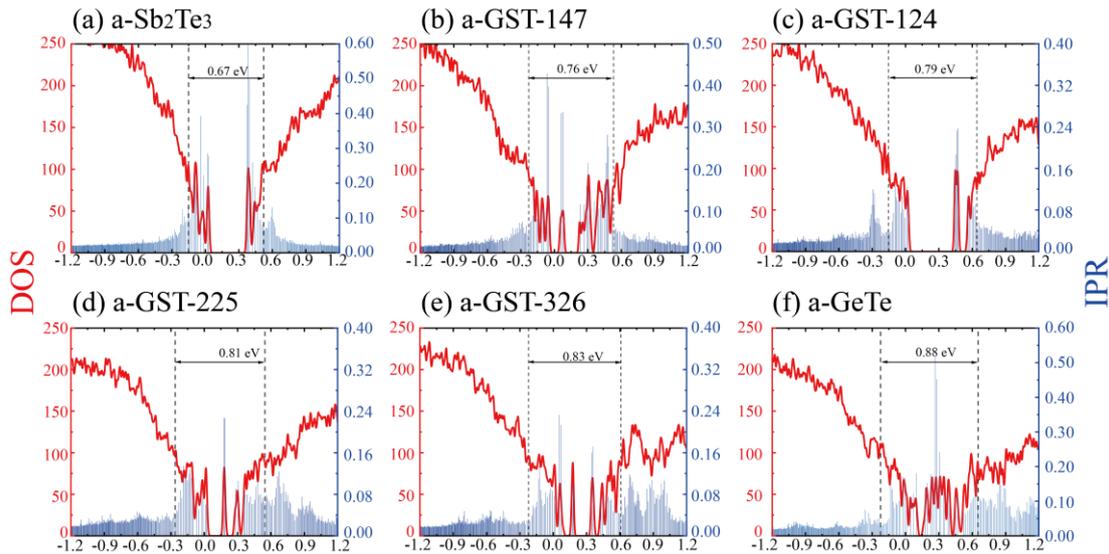

**Figure 7.** The density of states as well as the corresponding normalized IPR for various amorphous models, calculated using shGGA-1/2+SOC. (a) a-$Sb_2Te_3$; (b) a-GST-147; (c) a-GST-124; (d) a-GST-225; (e) a-GST-326; (d) a-GeTe. The Fermi level corresponds to zero energy.

Analysis of the CN in an amorphous structure is more challenging compared with crystalline phases. We first applied the traditional integration method using the radial distribution function, $RDF(r)$. The mathematical details of this method is given in **Supplementary Note 4**. Moreover, the CNs for the amorphous phases were also calculated within the MLAC context. **Table 6** shows the average number of atoms in each specific CN, for all six amorphous materials. The averaging was performed over 3000 AIMD steps, with a time interval of 2 fs. The MLAC theory gives the CN values slightly different from the traditional method, but the overall trend is consistent. The speed of the MLAC statistics is remarkable since it only involves calculating the bond lengths and bond angles.

**Table 6.** Number of atoms in a specific CN configuration per supercell, in several amorphous models. The value outside the parentheses was obtained through the MLAC method, while the value inside the parentheses was obtained through the tradition RDF integration method.

|  | | Number of atoms per supercell | | |
|---|---|---|---|---|
|  | CN | Ge | Sb | Te |
| GST-147 | 1 | 0.1 (0.0) | 1.8 (0.0) | 3.6 (0.6) |
|  | 2 | 0.4 (0.0) | 8.2 (0.5) | 16.4 (54.3) |
|  | 3 | 2.7 (3.3) | 11.8 (42.8) | 49.2 (83.5) |
|  | 4 | 12.5 (14.4) | 28.7 (37) | 55.3 (19.7) |
|  | 5 | 5.5 (4.6) | 27.8 (10.4) | 29.4 (2.8) |
|  | 6 | 1.6 (1.0) | 13.5 (1.4) | 7.0 (0.0) |
|  | 7 | 0.1 (0.0) | 0.2 (0.0) | 0.2 (0.0) |
|  | 8 | 0.0 (0.0) | 0.0 (0.0) | 0.0 (0.0) |
| GST-124 | CN | Ge | Sb | Te |
|  | 1 | 0.0 (0.0) | 0.3 (0.0) | 0.7 (0.3) |
|  | 2 | 0.2 (0.0) | 1.5 (0.0) | 13.2 (22.5) |
|  | 3 | 2.3 (13.4) | 10.9 (2.8) | 52.8 (71.2) |
|  | 4 | 17.7 (21) | 24.7 (17.6) | 61.2 (50.7) |
|  | 5 | 14.3 (4.6) | 29.2 (33.6) | 27.3 (13.3) |
|  | 6 | 5.5 (0.8) | 13.2 (22.5) | 4.7 (1.9) |
|  | 7 | 0.0 (0.0) | 0.1 (3.2) | 0.1 (0.0) |
|  | 8 | 0.0 (0.0) | 0.0 (0.0) | 0.0 (0.0) |
| GST-225 | CN | Ge | Sb | Te |
|  | 1 | 0.1 (0) | 0.1 (0.0) | 1.4 (0.1) |
|  | 2 | 0.5 (0) | 1.1 (0) | 15.2 (29.6) |
|  | 3 | 3.5 (5.1) | 7.5 (5.5) | 54.5 (76.5) |
|  | 4 | 30.3 (31.3) | 18.3 (19.6) | 54.8 (38.3) |
|  | 5 | 19.5 (18.5) | 22.1 (23.6) | 21.2 (5.3) |
|  | 6 | 6.1 (4.7) | 10.7 (11) | 2.8 (0.2) |

|  | CN | Ge | Sb | Te |
|---|---|---|---|---|
|  | 7 | 0.0 (0.5) | 0.1 (0.3) | 0.1 (0.0) |
|  | 8 | 0.0 (0.0) | 0.0 (0.0) | 0.0 (0.0) |
|  | CN | Ge | Sb | Te |
| GST-326 | 1 | 0.2 (0.0) | 0.2 (0.0) | 0.9 (0) |
|  | 2 | 0.8 (0.0) | 0.6 (0.0) | 10.3 (15.2) |
|  | 3 | 5.8 (6.1) | 7.1 (2.4) | 45.1 (68.4) |
|  | 4 | 37.7 (38.4) | 16.7 (12.4) | 62.6 (46.2) |
|  | 5 | 23.0 (24.5) | 16.4 (18.8) | 26.7 (15.7) |
|  | 6 | 7.6 (5.7) | 8.7 (15) | 4.3 (4) |
|  | 7 | 0.1 (0.2) | 0.2 (1.4) | 0.1 (0.3) |
|  | 8 | 0.0 (0.0) | 0.0 (0.0) | 0.0 (0.1) |
|  | CN | Ge | Sb | Te |
| Sb$_2$Te$_3$ | 1 | - | 0.4 (0) | 1.5 (0.2) |
|  | 2 | - | 2.1 (0) | 18.9 (24.8) |
|  | 3 | - | 12.7 (8.1) | 54.9 (84.1) |
|  | 4 | - | 32.9 (30) | 62.5 (54.2) |
|  | 5 | - | 42.8 (45) | 35.0 (15) |
|  | 6 | - | 28.8 (32.0) | 7.0 (1.7) |
|  | 7 | - | 0.4 (4.6) | 0.1 (0.1) |
|  | 8 | - | 0.0 (0.3) | 0.0 (0.0) |
|  | CN | Ge | Sb | Te |
| GeTe | 1 | 0.7 (0.0) | - | 2.5 (0.0) |
|  | 2 | 2.5 (0.0) | - | 11.2 (17.5) |
|  | 3 | 10.7 (20.2) | - | 53.3 (95.4) |
|  | 4 | 77.7 (86.1) | - | 61.9 (33) |
|  | 5 | 43.1 (36.2) | - | 18.9 (4) |
|  | 6 | 15.2 (7.3) | - | 2.2 (0.1) |
|  | 7 | 0.1 (0.1) | - | 0.0 (0.0) |
|  | 8 | 0.0 (0.0) | - | 0.0 (0.0) |

## V. Conclusion

We report two effective and efficient methods in analyzing the phase change materials exemplified by GST. The shell GGA-1/2 method, as a self-energy correction method at the LDA/GGA computational complexity, is shown to correct the electronic structures of crystalline Sb$_2$Te$_3$ (predicted band gap: 0.27 eV), GST-147 (0.45 eV), GST-124 (0.59 eV), GST-225 (0.57 eV) and GST-326 (0.61 eV) faithfully, especially in terms of the band gaps. In contrast, plain GGA calculation could yield acceptable band gaps only if the spin-orbit coupling effect is neglected, but the impact of spin-orbit coupling is distinct in the entire series of GST compounds, which inevitably renders inconsistent physical results. The calculated band gap of rhombohedral GeTe using the GGA-1/2 method with SOC correction yields a value of 1.31 eV. While this result slightly overestimates the experimental measurement, it demonstrates remarkable agreement with the more computationally demanding HSE06+SOC calculation (1.21 eV). Notably, despite achieving comparable accuracy to the HSE06 functional, the GGA-1/2 approach exhibits superior computational efficiency, with a speed enhancement of nearly three orders of magnitude. Shell DFT-1/2+SOC also predicts reasonable mobility gaps in amorphous GST (*e.g.*, a mobility gap of 0.81 eV for amorphous GST-225).

On the other hand, the mixed length-angle coordination theory has been shown to be suitable for analyzing the coordination numbers in amorphous GST. While giving exactly identical coordination numbers of crystalline GST as the traditional radial distribution function integration method, the new theory yields similar but not exactly the same results for amorphous GST samples. The new method is efficient in that it does not require any integration, but the definite coordination number of a specified atom could be straightforwardly obtained by inspecting the bond angles. This new perspective may afford a more effective coordination number analysis in the amorphous phase of phase change materials.

## Acknowledgement

This work was supported by the National Science and Technology Major Project of China (Grant No. 2022ZD0117600) and the National Natural Science Foundation of China under Grant No. 12474230.

Supplementary Information for

# High-efficiency computational methodologies for electronic properties and structural characterization of Ge-Sb-Te based phase change materials

Shanzhong Xie, Kan-Hao Xue,* Shaojie Yuan, Shengxin Yang, Heng Yu, Rongchuan Gu, Ming Xu,* Xiangshui Miao

School of Integrated Circuits, Huazhong University of Science and Technology, Wuhan 430074, China

*Corresponding authors, email: xkh@hust.edu.cn (K.-H. X.), mxu@hust.edu.cn (M. X.).

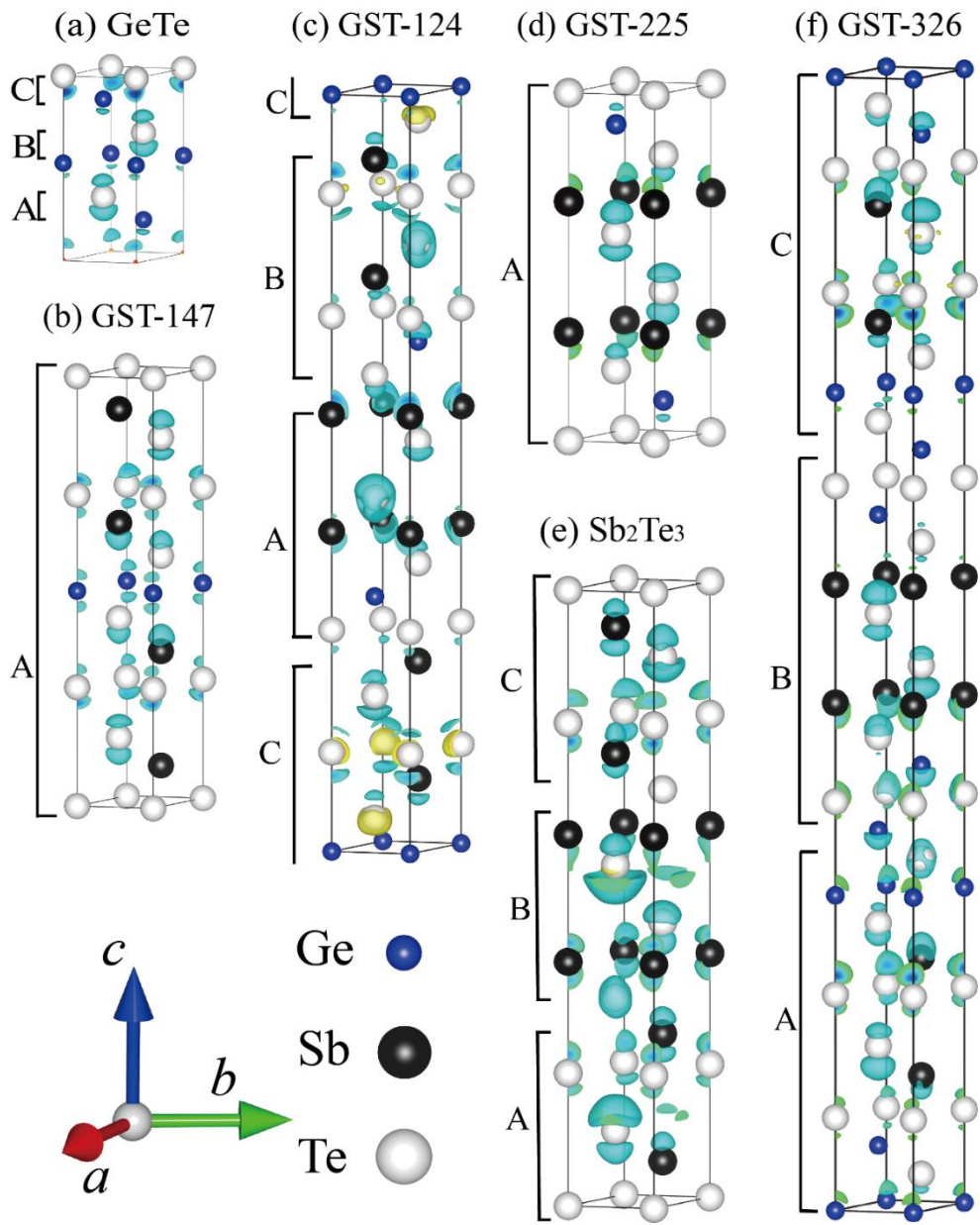

**Figure S1**. Iso-surface charts of the charge density difference for (a) GeTe (contour density $14.85 \times 10^{-3}$ Å$^{-3}$); (b) GST-147 (contour density $8.10 \times 10^{-3}$ Å$^{-3}$); (c) GST-124 (contour density $4.93 \times 10^{-3}$ Å$^{-3}$); (d) GST-225 (contour density $13.50 \times 10^{-3}$ Å$^{-3}$); (e) Sb$_2$Te$_3$ (contour density $5.06 \times 10^{-3}$ Å$^{-3}$); (f) GST-326 (contour density $4.45 \times 10^{-3}$ Å$^{-3}$).

# Supplementary Note 1. Detailed $k$ point settings in band structure calculations as well as the computational speed tests

In this work, an electronic band structure calculation at the GGA or shGGA-1/2 level involves two steps, a self-consistent one and a non-self-consistent one. The first step involves an equal-spacing Monkhorst-Pack $k$-mesh setting. We have always used fine k-meshes centered at the Γ point for the self-consistent runs. In the non-self-consistent step, the charge density profile is fixed as the output of the previous run. Meanwhile, the $k$ points only involve those distributed along certain selected lines in the first Brillouin zone. The benefit of this two-step calculation lies in the possible extremely fine $k$ point resolution in the second step.

For HSE06 calculations, we have to obtain the band structures in a self-consistent manner. This involves listing the Monkhorst-Pack $k$ mesh, including their weights, together with some zero-weight $k$ points. The zero-weight $k$ points cover those pre-selected $k$-lines. In the following table, detailed $k$ point settings are shown, including the Monkhorst-Pack part as well as the line-mode k point setting. Nevertheless, it has to be emphasized that, for GGA and shGGA-1/2 calculations, the Monkhorst-Pack mesh is only relevant to the first self-consistent step, while the line-mode setting is only relevant to the second non-self-consistent step. For HSE06 data, however, both Monkhorst-Pack and line-mode settings yield the single $k$ point setting for the self-consistent calculation.

Table S1. Detailed $k$-point settings in the actual band structure calculations

|  |  | $k$ point setting | | | | | |
|---|---|---|---|---|---|---|---|
|  |  | GGA | GGA+SOC | shGGA-1/2 | shGGA-1/2+SOC | HSE06 | HSE06+SOC |
| Sb$_2$Te$_3$ | Monkhorst-Pack: 21×21×5 | 21×21×5 | 21×21×5 | 21×21×5 | 17×17×5 | 11×11×3 |
|  | K points per line: 301 | 301 | 301 | 301 | 147 | 24 |
| GST-147 | | 21×21×5 | 21×21×5 | 21×21×5 | 21×21×5 | 15×15×3 | 11×11×3 |
|  | | 301 | 301 | 301 | 301 | 73 | 21 |
| GST-124 | | 21×21×5 | 21×21×5 | 21×21×5 | 21×21×5 | 15×15×5 | 10×10×1 |
|  | | 301 | 301 | 301 | 301 | 119 | 26 |
| GST-225 | | 21×21×5 | 21×21×5 | 21×21×5 | 21×21×5 | 17×17×5 | 11×11×3 |
|  | | 301 | 301 | 301 | 301 | 147 | 20 |
| GST- | | 21×21×5 | 21×21×5 | 21×21×5 | 21×21×5 | 15×15×1 | 10×10×1 |

| 326 | | | | | | |
|---|---|---|---|---|---|---|
| | 301 | 301 | 301 | 301 | 27 | 26 |
| GeTe | 21×21×5 | 21×21×5 | 21×21×5 | 21×21×5 | 21×21×7 | 11×11×3 |
| | 301 | 301 | 301 | 301 | 151 | 72 |

Although the settings above could guarantee the best accuracy with computational load still tolerable, it poses a difficulty in comparing the efficiency of shGGA-1/2 compared with HSE06. This is partly because shGGA-1/2 allows for much more irreducible $k$ points. And it is also partly because shGGA-1/2 involves a two-step calculation, while that of HSE06 is merely one-step. To make a fair comparison, we re-did the GGA and shGGA-1/2 calculations (either with or without SOC) in the self-consistent manner, with exactly the same $k$ point setting as that of the HSE06 calculation. The test results are demonstrated below.

Table S2. Total time costs of the calculations

| | Time (second) | | | | | |
|---|---|---|---|---|---|---|
| | GGA | GGA+SOC | shGGA-1/2 | shGGA-1/2+SOC | HSE06 | HSE06+SOC |
| $Sb_2Te_3$ | 167 | 527 | 161 | 584 | 1185018 (~14 d) | 3736252 (~43 d) |
| GST-147 | 161 | 333 | 146 | 272 | 892,481 (~11 d) | 4604801 (~53 d) |
| GST-124 | 294 | 552 | 293 | 548 | 1,803,972 (~21 d) | 604356 (~7 d) |
| GST-225 | 50 | 121 | 43 | 148 | 864186 (~10 d) | 918017 (~11 d) |
| GST-326 | 83 | 1555 | 81 | 1726 | 596,809 (~7 d) | 3600014 (~41 d) |
| GeTe | 24 | 99 | 25 | 111 | 329391 (~4 d) | 424517 (~5 d) |

Table S3. The number of irreducible k-points when testing calculations

| | GGA | GGA+SOC | shGGA-1/2 | shGGA-1/2+SOC | HSE06 | HSE06+SOC |
|---|---|---|---|---|---|---|
| $Sb_2Te_3$ | 147 | 182 | 147 | 182 | 147 | 182 |
| GST-147 | 73 | 182 | 73 | 182 | 73 | 182 |
| GST-124 | 119 | 52 | 119 | 52 | 119 | 52 |
| GST-225 | 147 | 182 | 147 | 182 | 147 | 182 |
| GST-326 | 27 | 52 | 27 | 52 | 27 | 52 |
| GeTe | 303 | 363 | 303 | 363 | 303 | 363 |

## Supplementary Note 2. The mixed length-angle coordination number theory[1]

The mixed length-angle coordination (MLAC) theory could be used to evaluate the coordination number (CN) for an atom/ion in a bulk solid. Since even one element may involve several types of atoms, each with a different CN, the CN should be assigned for a particular type of atomic site, rather than broadly for an element.

Suppose one needs to find the CN for ion A in a given solid. The MLAC procedure is as follows. We take an amorphous GST model as the example, where A corresponds to Ge.

(i) Select a central A ion, and find out the neighboring ions that could chemically form a bond with A. Make a set of B={$B_0$, $B_1$, $B_2$, …} according to their distance to A, in an ascending order. In other words, $B_0$ is the closest counter ion from A, and $B_1$ is the second closest. In our example, we chose a central Ge atom in a-GST-124. The set B is {$Te_0$ (bond length $L_{Ge-Te_0}$ = 2.57 Å), $Te_1$ ($L_{Ge-Te_1}$ = 2.65 Å), $Te_2$ ($L_{Ge-Te_2}$ = 2.67 Å), $Te_3$ ($L_{Ge-Te_3}$ = 2.73 Å), Ge' ($L_{Ge-Ge'}$ = 3.55 Å), ……}.

(ii) The closest ion $B_0$ is automatically counted within the coordination. In this example, $Te_0$ is automatically counted as a neighbor of the central Ge atom, within its coordination.

(iii) For $B_1$, one calculates the bond angle $\angle B_0AB_1$. If it is greater than the critical value $\theta_{th}$ = 65°, then $B_1$ is within the coordination. Usually this is the case. In this example, for $B_1$ = $Te_1$ with a distance of 2.65 Å from the central Ge atom, $\angle Te_0$-$Ge_1$-$Te_1$=107.7° > 65°. Hence, $Te_1$ is within the coordination.

(iv) For $B_2$, one has to calculate two bond angles $\angle B_0AB_2$ and $\angle B_1AB_2$., i.e., considering the two ions already within the coordination $B_0$ and $B_1$. In case both angles are greater than $\theta_{th}$, then $B_2$ is within the coordination. Otherwise, the counting finishes, and the CN for A is 2. In the concrete example, $B_2$ = $Te_2$, which is 2.67 Å apart from the central Ge atom. One has to calculate two angles because there are already two atoms within the coordination. The results are $\angle Te_0$-$Ge_1$-$Te_2$ = 109.4° and $\angle Te_1$-$Ge_1$-$Te_2$ = 111.5°, both greater than 65°. Hence, $Te_2$ is considered to be within the coordination.

(v) For $B_3$, one has to calculate three bond angles $\angle B_0AB_3$, $\angle B_1AB_3$ and $\angle B_2AB_3$. On condition that all the three angles are greater than $\theta_{th}$, then $B_3$ is within the coordination. Otherwise, the counting finishes, and the CN for A is 3. In the example, for $B_3$ = $Te_3$, one obtains $\angle Te_0$-$Ge_1$-$Te_3$ = 113.9°, $\angle Te_1$-$Ge_1$-$Te_3$ = 99.9°, and $\angle Te_2$-$Ge_1$-$Te_3$ = 114.1°. Since all three angles are greater than 65°, $Te_3$ is within the coordination of the central Ge atom.

(vi) The process further continues. For $B_n$, one should calculate the $n$ angles formed between A—$B_n$ and A—$B_0$, A—$B_1$, …, A—$B_{n-1}$. In case any of the angle is lower than $\theta_{th}$, then the counting finishes, and the CN for A is $n$. If, otherwise, all these angles are greater than $\theta_{th}$, then one should start to examine the possibility to include $B_{n+1}$ in the coordination. In our example, we process as follows. For $B_4$ = Ge' that is 3.55 Å apart from the central Ge atom, four angles are calculated as $\angle Te_0$-$Ge_1$-$Ge_4$ = 54.5°, $\angle Te_1$-$Ge_1$-$Ge_4$ = 54.9°, $\angle Te_2$-$Ge_1$-$Ge_4$ = 114.4°, and $\angle Te_3$-$Ge_1$-$Ge_4$ = 130.9°. Among them, at least one angle is smaller than 65°. Hence, $B_4$ = Ge' is outside the set of coordination, and the CN of the central Ge atom is fixed to be 4.

(vii) There are several critical angles involved in the counting, $\theta_1$, $\theta_2$ as well as $\theta_{th}$. $\theta_1$ is defined as the minimum angle value among $\angle B_iAB_j$, where $B_i$ and $B_j$ 为 are any two distinct counter ions within the coordination. On the other hand, $\theta_2$ is defined as the maximum angle value among $\angle B_iAB_n$, where $B_i$ is an arbitrary counter ion within the coordination, and $B_n$ is the first counter ion outside the coordination. Following the origin work, $\theta_{th}$ is selected as 65°. It is logically guaranteed that $\theta_2 \leqslant \theta_{th} \leqslant \theta_1$.

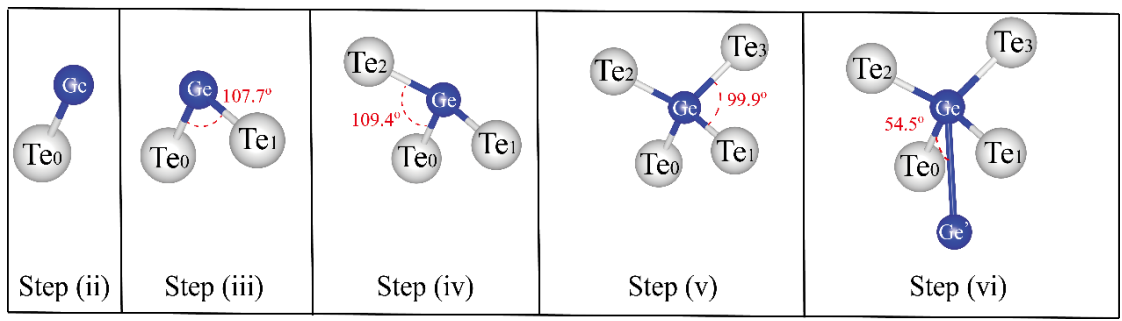

**Figure S2**. An example in counting the CN for a selected Ge atom in a-GST-124.

## Supplementary Note 3. Detailed coordination analyses for Sb$_2$Te$_3$, GST and GeTe

In **Table S4**, the general features regarding the coordination environment in the six compounds are analyzed. **Figures S3一S8** illustrate the detailed coordination environment information in the six compounds, respectively.

Based on the MLAC theory, the coordination number for each atom could be figured out unambiguously. The required bond lengths ($L$, in Å) and bond angles ($\theta$) are given in **Table S5** for the six compounds.

Furthermore, we extracted two special angles ($\theta_1$ and $\theta_2$) that were finally used to determine the coordination numbers, as shown in **Table S6**.

**Table S4.** Chemical environments in the six compounds

| Material | Coordination environment | | Global composition |
|---|---|---|---|
| | Metal site | Te site | |
| GST-147 | Ge (VI) | Te1 (VI): bonding with Sb only | 1 Ge (VI) + 4 Sb (VI) + 1 |
| | Sb (VI) | Te2 (VI): bonding with Ge and Sb | Te1 (VI) + 2 Te2 (VI) + 4 |
| | | Te3 (III): bonding with Sb only | Te3 (III) |
| GST-124 | Ge (VI) | Te2 (VI): bonding with Ge and Sb | 3 Ge (VI) + 6 Sb (VI) + 6 |
| | Sb (VI) | Te3 (III): bonding with Sb only | Te2 (VI) + 6 Te3 (III) |
| GST-225 | Ge (VI) | Te1 (VI): bonding with Ge only | 2 Ge (VI) + 2 Sb (VI) + 1 |
| | Sb (VI) | Te2 (VI): bonding with Ge and Sb | Te1 (VI) + 2 Te2 (VI) + 2 |
| | | Te3 (III): bonding with Sb only | Te3 (III) |
| GST-326 | Ge (VI) | Te1 (VI): bonding with Ge only | 9 Ge (VI) + 6 Sb (VI) + 6 |
| | Sb (VI) | Te2 (VI): bonding with Ge and Sb | Te1 (VI) + 6 Te2 (VI) + 6 |
| | | Te3 (III): bonding with Sb only | Te3 (III) |
| Sb$_2$Te$_3$ | Sb (VI) | Te1 (VI): bonding with Sb only | 6 Sb (VI) + 3 Te1 (VI) + 6 |
| | | Te3 (III): bonding with Sb only | Te3 (III) |
| GeTe | Ge (VI) | Te1 (VI): bonding with Ge only | 3 Ge (VI) + 3 Te (VI) |

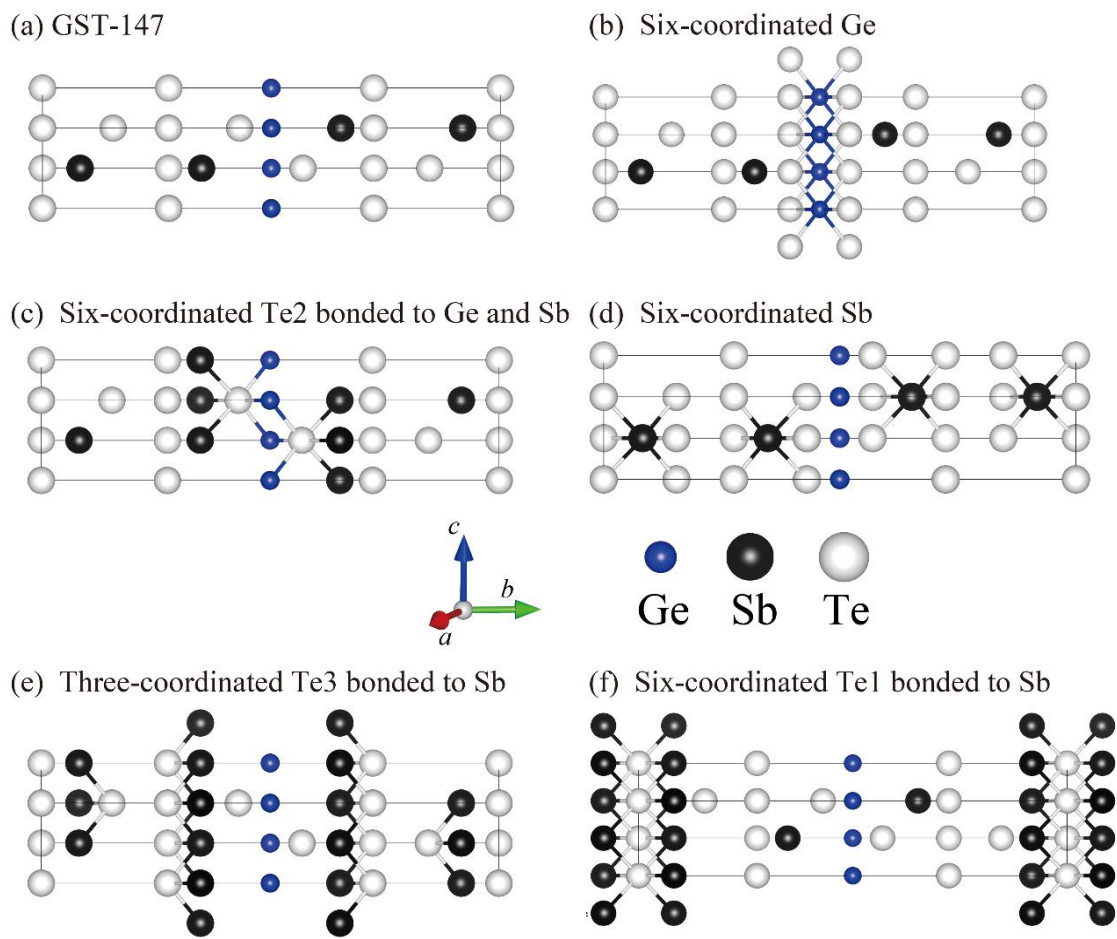

**Figure S3.** (a) A schematic crystal structure of GST-147, which comprises the following characteristic coordination units. (b) One six-coordinated (6C) Ge atom. (c) Two 6C Te2 atoms bonded to both Ge and Sb. (d) Four 6C Sb atoms. (e) Four 3C Te3 atoms bonded solely to Sb. (f) One 6C Te1 atom bonded exclusively to Sb.

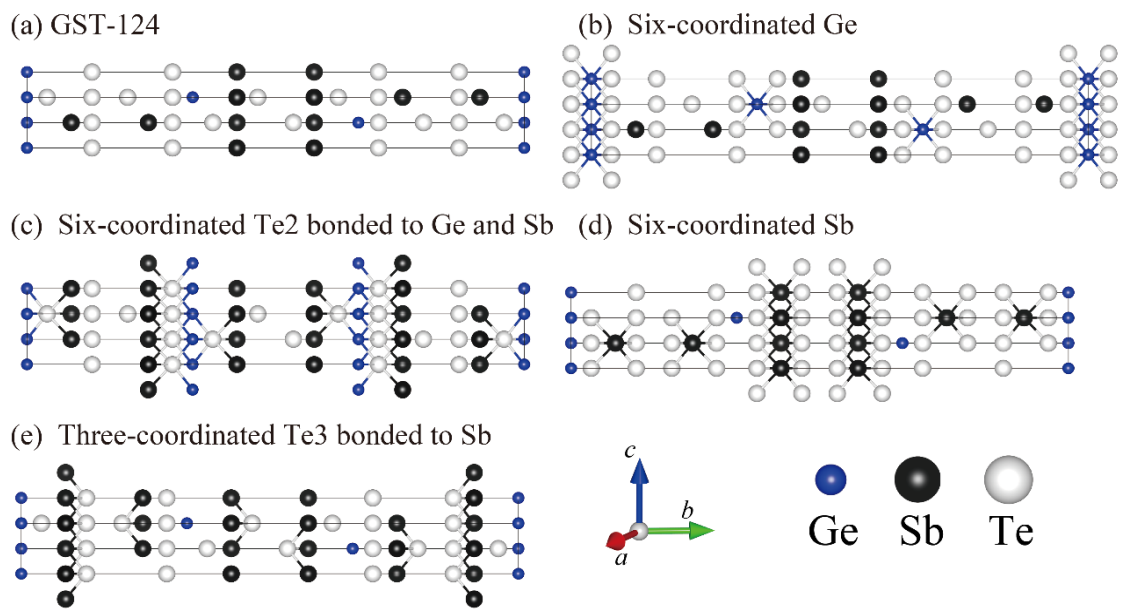

**Figure S4.** (a) A schematic crystal structure of GST-124, which comprises the following characteristic coordination units. (b) Three 6C Ge atom. (c) Six 6C Te2 atoms bonded to both Ge and Sb. (d) Six 6C Sb atoms. (e) Six 3C Te3 atom bonded exclusively to Sb.

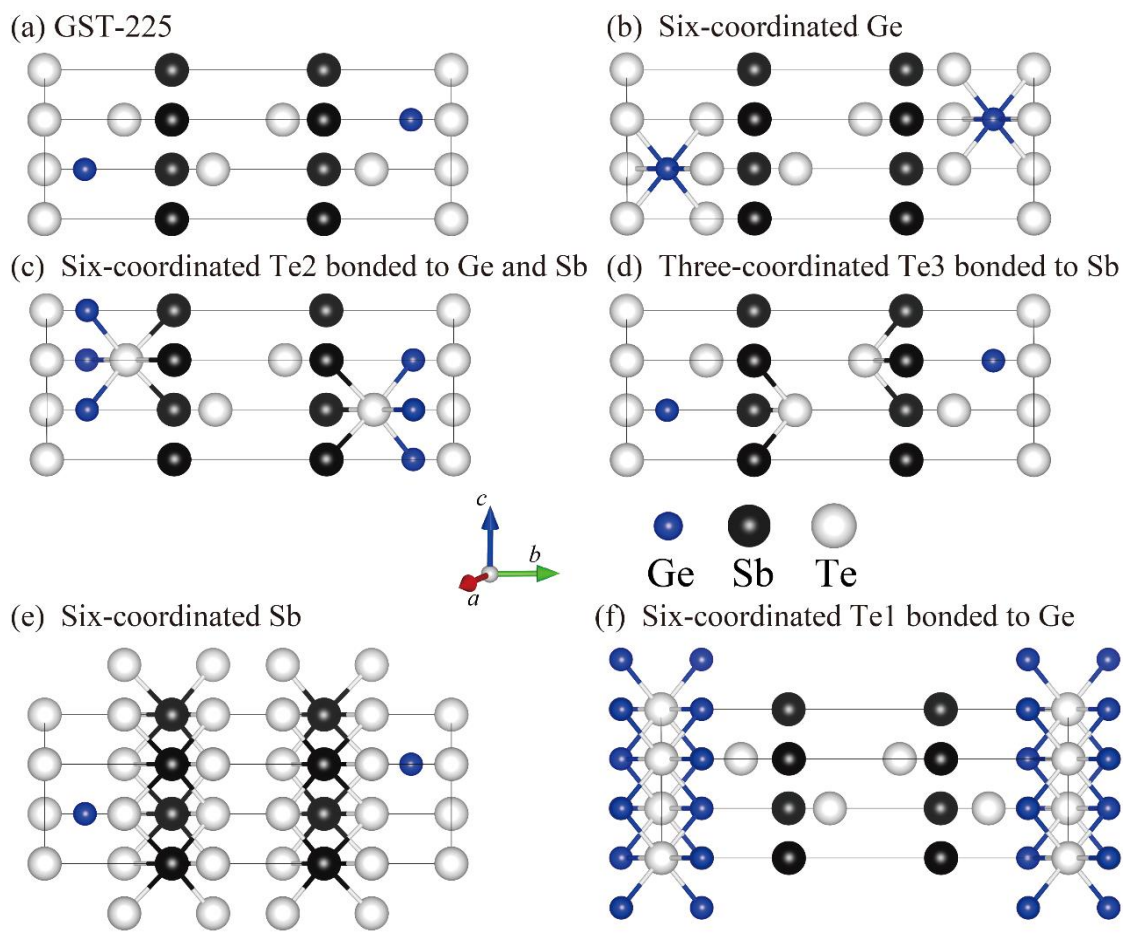

**Figure S5.** (a) A schematic crystal structure of GST-225, which comprises the following characteristic coordination units. (b) Two 6C Ge atom. (c) Two 6C Te2 atoms bonded to both Ge and Sb. (d) Two 3C Te3 atoms bonded solely to Sb. (e) Two 6C Sb atoms. (f) One 6C Te1 atom bonded exclusively to Ge.

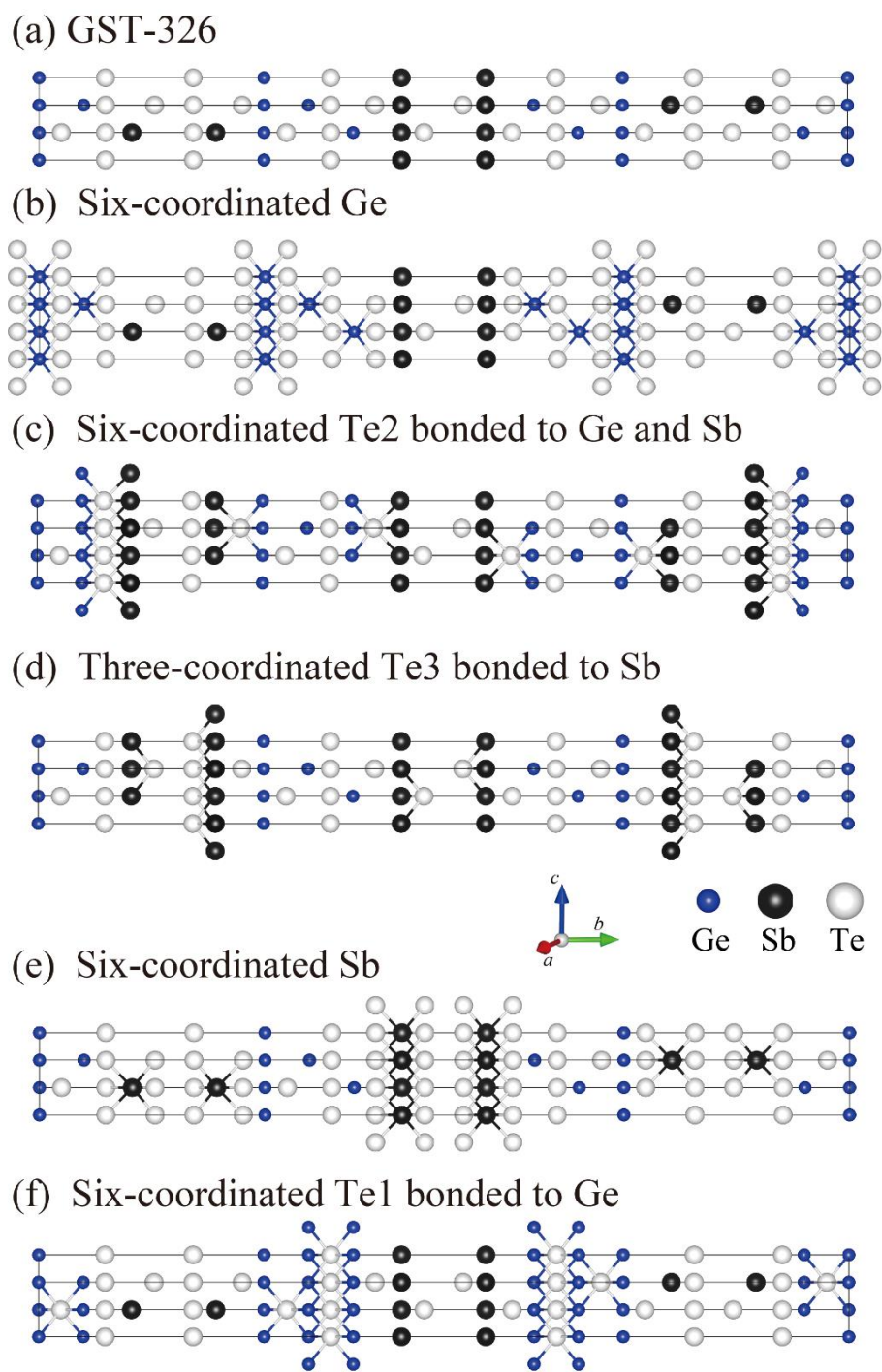

**Figure S6.** (a) A schematic crystal structure of GST-326, which comprises the following characteristic coordination units. (b) Nine 6C Ge atom. (c) Six 6C Te2 atoms bonded to both Ge and Sb. (d) Six 3C Te3 atoms bonded solely to Sb. (e) Six 6C Sb atoms. (f) Six 6C Te1 atom bonded exclusively to Ge.

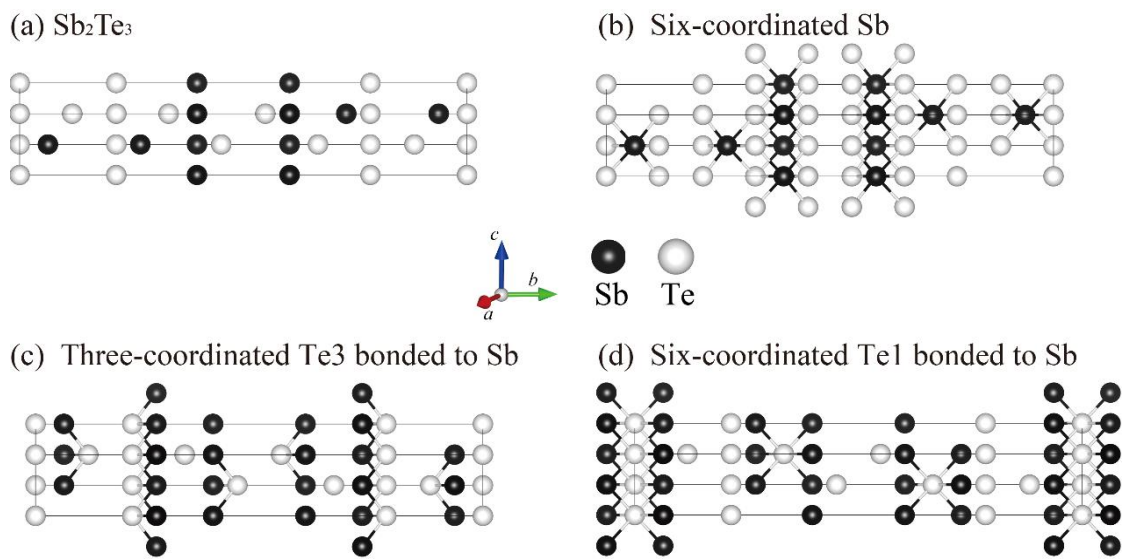

**Figure S7.** (a) A schematic crystal structure of $Sb_2Te_3$, which comprises the following characteristic coordination units. (b) Six 6C Sb atoms. (c) Six 3C Te3 atoms bonded solely to Sb. (d) Three 6C Te1 atom bonded exclusively to Sb.

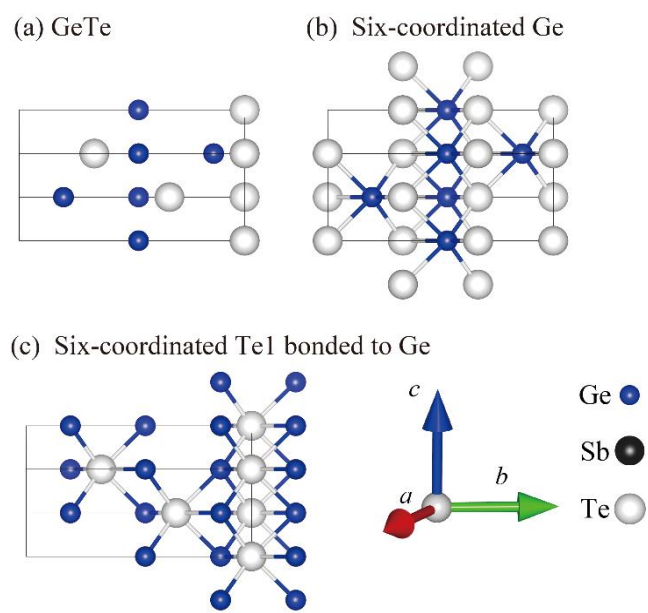

**Figure S8.** (a) A schematic crystal structure of GeTe, which comprises the following characteristic coordination units. (b) Three 6C Ge atom. (c) Three 6C Te1 atom bonded exclusively to Ge.

**Table S5.** Statistics of the bond length $L$ (Å) and bond angle $\theta$ (°) in the six compounds. Blue and red angle values represent $\theta_1$ and $\theta_2$, respectively.

| | Ge, CN=6 | | | | Sb, CN=6 | | | |
|---|---|---|---|---|---|---|---|---|
| | Bond | $L$ | Bond | $\theta$ | Bond | $L$ | Bond | $\theta$ |
| GST-147 | Ge-Te1 | 2.96 | - | - | Sb-Te1 | 2.99 | - | - |
| | Ge-Te2 | 2.96 | Te1-Ge-Te2 | 91.3 | Sb-Te2 | 2.99 | Te1-Sb-Te2 | 90.0 |
| | Ge-Te3 | 2.96 | Te1-Ge-Te3 | 91.3 | Sb-Te3 | 2.99 | Te1-Sb-Te3 | 90.0 |
| | | | Te2-Ge-Te3 | 91.3 | | | Te2-Sb-Te3 | 90.0 |
| | Ge-Te4 | 2.96 | Te1-Ge-Te4 | 88.7 | Sb-Te4 | 3.15 | Te1-Sb-Te4 | 92.6 |
| | | | Te2-Ge-Te4 | 88.7 | | | Te2-Sb-Te4 | 92.6 |
| | | | Te3-Ge-Te4 | 180 | | | Te3-Sb-Te4 | 176.3 |
| | Ge-Te5 | 2.96 | Te1-Ge-Te5 | 88.7 | Sb-Te5 | 3.15 | Te1-Sb-Te5 | 92.6 |
| | | | Te2-Ge-Te5 | 180 | | | Te2-Sb-Te5 | 176.3 |
| | | | Te3-Ge-Te5 | 88.7 | | | Te3-Sb-Te5 | 92.6 |
| | | | Te4-Ge-Te5 | 91.3 | | | Te4-Sb-Te5 | 84.7 |
| | Ge-Te6 | 2.96 | Te1-Ge-Te6 | 180 | Sb-Te6 | 3.15 | Te1-Sb-Te6 | 176.3 |
| | | | Te2-Ge-Te6 | **88.7** | | | Te2-Sb-Te6 | 92.6 |
| | | | Te3-Ge-Te6 | **88.7** | | | Te3-Sb-Te6 | 92.6 |
| | | | Te4-Ge-Te6 | 91.3 | | | Te4-Sb-Te6 | **84.7** |
| | | | Te5-Ge-Te6 | 91.3 | | | Te5-Sb-Te6 | **84.7** |
| | Ge-Te7 | 5.16 | Te1-Ge-Te7 | 55.0 | Sb-Te7 | 5.18 | Te1-Sb-Te7 | **54.7** |
| | | | Te2-Ge-Te7 | 126.8 | | | Te2-Sb-Te7 | **54.7** |
| | | | Te3-Ge-Te7 | 55.0 | | | Te3-Sb-Te7 | 125.3 |
| | | | Te4-Ge-Te7 | 125.0 | | | Te4-Sb-Te7 | 58.4 |
| | | | Te5-Ge-Te7 | **53.2** | | | Te5-Sb-Te7 | 125.2 |
| | | | Te6-Ge-Te7 | 125.0 | | | Te6-Sb-Te7 | 125.2 |
| | Te1, CN=6, bonded to Sb | | | | Te2, CN=6, bonded to Ge and Sb | | | |
| | Bond | $L$ | Bond | $\theta$ | Bond | $L$ | Bond | $\theta$ |
| | Te1-Sb1 | 3.14 | - | - | Te2-Ge1 | 2.96 | - | - |
| | Te1-Sb2 | 3.14 | Sb1-Te1-Sb2 | 84.7 | Te2-Ge2 | 2.96 | Ge1-Te2-Ge2 | 91.3 |
| | Te1-Sb3 | 3.14 | Sb1-Te1-Sb3 | 84.7 | Te2-Ge3 | 2.96 | Ge1-Te2-Ge3 | 91.3 |
| | | | Sb2-Te1-Sb3 | 95.3 | | | Ge2-Te2-Ge3 | 91.3 |
| | Te1-Sb4 | 3.14 | Sb1-Te1-Sb4 | 95.3 | Te2-Sb1 | 3.15 | Ge1-Te2-Sb1 | 175.1 |
| | | | Sb2-Te1-Sb4 | 95.3 | | | Ge2-Te2-Sb1 | 92.1 |
| | | | Sb3-Te1-Sb4 | 180 | | | Ge3-Te2-Sb1 | 92.1 |
| | Te1-Sb5 | 3.14 | Sb1-Te1-Sb5 | 95.3 | Te2-Sb2 | 3.15 | Ge1-Te2-Sb2 | 92.1 |
| | | | Sb2-Te1-Sb5 | 180 | | | Ge2-Te2-Sb2 | 92.1 |
| | | | Sb3-Te1-Sb5 | 95.3 | | | Ge3-Te2-Sb2 | 175.1 |
| | | | Sb4-Te1-Sb5 | 84.7 | | | Sb1-Te2-Sb2 | 84.3 |
| | Te1-Sb6 | 3.14 | Sb1-Te1-Sb6 | 180 | Te2-Sb3 | 3.15 | Ge1-Te2-Sb3 | 92.1 |
| | | | Sb2-Te1-Sb6 | 95.3 | | | Ge2-Te2-Sb3 | 175.1 |
| | | | Sb3-Te1-Sb6 | 95.3 | | | Ge3-Te2-Sb3 | 92.1 |

|  |  | Bond | θ |  |  | Bond | θ |
|---|---|---|---|---|---|---|---|
|  |  | Sb4-Te1-Sb6 | **84.7** |  |  | Sb1-Te2-Sb3 | **84.3** |
|  |  | Sb5-Te1-Sb6 | **84.7** |  |  | Sb2-Te2-Sb3 | **84.3** |
|  | Te1-Sb7 | Sb1-Te1-Sb7 | **53.4** |  | Te2-Ge4 | Ge1-Te2-Ge4 | **55.0** |
|  |  | Sb2-Te1-Sb7 | 119.0 |  |  | Ge2-Te2-Ge4 | 126.8 |
|  | 5.27 | Sb3-Te1-Sb7 | **53.4** |  | 5.16 | Ge3-Te2-Ge4 | **55.0** |
|  |  | Sb4-Te1-Sb7 | 125.6 |  |  | Sb1-Te2-Ge4 | 124.8 |
|  |  | Sb5-Te1-Sb7 | 61.0 |  |  | Sb2-Te2-Ge4 | 124.8 |
|  |  | Sb6-Te1-Sb7 | 126.6 |  |  | Sb3-Te2-Ge4 | 58.1 |

| Te3, CN=3, bonded to Sb |  |  |  |
|---|---|---|---|
| Bond | L | Bond | θ |
| Te3-Sb1 | 2.99 | - | - |
| Te3-Sb2 | 2.99 | Sb1-Te3-Sb2 | **90** |
| Te3-Sb3 | 2.99 | Sb1-Te3-Sb3 | **90** |
|  |  | Sb2-Te3-Sb3 | **90** |
| Te3-Sb4 | 5.18 | Sb1-Te3-Sb4 | **54.8** |
|  |  | Sb2-Te3-Sb4 | 125.4 |
|  |  | Sb3-Te3-Sb4 | **54.8** |

GST-124

| Ge, CN=6 |  |  |  | Sb, CN=6 |  |  |  |
|---|---|---|---|---|---|---|---|
| Bond | L | Bond | θ | Bond | L | Bond | θ |
| Ge-Te1 | 2.96 | - | - | Sb-Te1 | 2.99 | - | - |
| Ge-Te2 | 2.96 | Te1-Ge-Te2 | 91.2 | Sb-Te2 | 2.99 | Te1-Sb-Te2 | 90 |
| Ge-Te3 | 2.96 | Te1-Ge-Te3 | 91.2 | Sb-Te3 | 2.99 | Te1-Sb-Te3 | 90 |
|  |  | Te2-Ge-Te3 | 91.2 |  |  | Te2-Sb-Te3 | 90 |
| Ge-Te4 | 2.96 | Te1-Ge-Te4 | 88.8 | Sb-Te4 | 3.15 | Te1-Sb-Te4 | 176.0 |
|  |  | Te2-Ge-Te4 | 88.8 |  |  | Te2-Sb-Te4 | 93.0 |
|  |  | Te3-Ge-Te4 | 180.0 |  |  | Te3-Sb-Te4 | 93.0 |
| Ge-Te5 | 2.96 | Te1-Ge-Te5 | 180.0 | Sb-Te5 | 3.15 | Te1-Sb-Te5 | 93.0 |
|  |  | Te2-Ge-Te5 | 88.8 |  |  | Te2-Sb-Te5 | 93.0 |
|  |  | Te3-Ge-Te5 | 88.8 |  |  | Te3-Sb-Te5 | 176.0 |
|  |  | Te4-Ge-Te5 | 91.2 |  |  | Te4-Sb-Te5 | 84.1 |
| Ge-Te6 | 2.96 | Te1-Ge-Te6 | **88.8** | Sb-Te6 | 3.15 | Te1-Sb-Te6 | 93.0 |
|  |  | Te2-Ge-Te6 | 180 |  |  | Te2-Sb-Te6 | 176.0 |
|  |  | Te3-Ge-Te6 | **88.8** |  |  | Te3-Sb-Te6 | 93.0 |
|  |  | Te4-Ge-Te6 | 91.2 |  |  | Te4-Sb-Te6 | **84.1** |
|  |  | Te5-Ge-Te6 | 91.2 |  |  | Te5-Sb-Te6 | **84.1** |
| Ge-Te7 | 5.16 | Te1-Ge-Te7 | 55.0 | Sb-Te7 | 5.17 | Te1-Sb-Te7 | **54.7** |
|  |  | Te2-Ge-Te7 | 55.0 |  |  | Te2-Sb-Te7 | 125.2 |
|  |  | Te3-Ge-Te7 | 126.7 |  |  | Te3-Sb-Te7 | **54.7** |
|  |  | Te4-Ge-Te7 | **53.3** |  |  | Te4-Sb-Te7 | 125.2 |
|  |  | Te5-Ge-Te7 | 125.0 |  |  | Te5-Sb-Te7 | 125.2 |
|  |  | Te6-Ge-Te7 | 125.0 |  |  | Te6-Sb-Te7 | 58.9 |

| Te2, CN=6, bonded to Ge and Sb |  |  |  | Te3, CN=3, bonded to Sb |  |  |  |
|---|---|---|---|---|---|---|---|
| Bond | L | Bond | θ | Bond | L | Bond | θ |

|  | Te2-Ge1 | 2.96 | - | - | Te3-Sb1 | 2.99 | - | - |
|---|---|---|---|---|---|---|---|---|
|  | Te2-Ge2 | 2.96 | Ge1-Te2-Ge2 | 91.2 | Te3-Sb2 | 2.99 | Sb1-Te3-Sb2 | 89.9 |
|  | Te2-Ge3 | 2.96 | Ge1-Te2-Ge3 | 91.2 | Te3-Sb3 | 2.99 | Sb1-Te3-Sb3 | **89.9** |
|  |  |  | Ge2-Te2-Ge3 | 91.2 |  |  | Sb2-Te3-Sb3 | **89.9** |
|  | Te2-Sb1 | 3.15 | Ge1-Te2-Sb1 | 92.3 | Te3-Sb4 | 5.18 | Sb1-Te3-Sb4 | 125.2 |
|  |  |  | Ge2-Te2-Sb1 | 92.3 |  |  | Sb2-Te3-Sb4 | **54.7** |
|  |  |  | Ge3-Te2-Sb1 | 175.1 |  |  | Sb3-Te3-Sb4 | **54.7** |
|  | Te2-Sb2 | 3.15 | Ge1-Te2-Sb2 | 92.3 | - | - | - | - |
|  |  |  | Ge2-Te2-Sb2 | 175.1 |  |  | - | - |
|  |  |  | Ge3-Te2-Sb2 | 92.3 |  |  | - | - |
|  |  |  | Sb1-Te2-Sb2 | 84.1 |  |  | - | - |
|  | Te2-Sb3 | 3.15 | Ge1-Te2-Sb3 | 175.1 | - | - | - | - |
|  |  |  | Ge2-Te2-Sb3 | 92.3 |  |  | - | - |
|  |  |  | Ge3-Te2-Sb3 | 92.3 |  |  | - | - |
|  |  |  | Sb1-Te2-Sb3 | **84.1** |  |  | - | - |
|  |  |  | Sb2-Te2-Sb3 | **84.1** |  |  | - | - |
|  | Te2-Ge4 | 5.16 | Ge1-Te2-Ge4 | **55.0** | - | - | - | - |
|  |  |  | Ge2-Te2-Ge4 | **55.0** |  |  | - | - |
|  |  |  | Ge3-Te2-Ge4 | 126.7 |  |  | - | - |
|  |  |  | Sb1-Te2-Ge4 | 58.3 |  |  | - | - |
|  |  |  | Sb2-Te2-Ge4 | 124.8 |  |  | - | - |
|  |  |  | Sb3-Te2-Ge4 | 124.8 |  |  | - | - |
| GST-225 | Ge, CN=6 | | | | Sb, CN=6 | | | |
|  | Bond | $L$ | Bond | $\theta$ | Bond | $L$ | Bond | $\theta$ |
|  | Ge-Te1 | 2.95 | - | - | Sb-Te1 | 2.99 | - | - |
|  | Ge-Te2 | 2.95 | Te1-Ge-Te2 | 91.3 | Sb-Te2 | 2.99 | Te1-Sb-Te2 | 89.7 |
|  | Ge-Te3 | 2.95 | Te1-Ge-Te3 | 91.3 | Sb-Te3 | 2.99 | Te1-Sb-Te3 | 89.7 |
|  |  |  | Te2-Ge-Te3 | 91.3 |  |  | Te2-Sb-Te3 | 89.7 |
|  | Ge-Te4 | 2.97 | Te1-Ge-Te4 | 89.1 | Sb-Te4 | 3.16 | Te1-Sb-Te4 | 175.9 |
|  |  |  | Te2-Ge-Te4 | 179.4 |  |  | Te2-Sb-Te4 | 93.2 |
|  |  |  | Te3-Ge-Te4 | 89.1 |  |  | Te3-Sb-Te4 | 93.2 |
|  | Ge-Te5 | 2.97 | Te1-Ge-Te5 | 179.4 | Sb-Te5 | 3.16 | Te1-Sb-Te5 | 93.2 |
|  |  |  | Te2-Ge-Te5 | 89.2 |  |  | Te2-Sb-Te5 | 175.9 |
|  |  |  | Te3-Ge-Te5 | 89.1 |  |  | Te3-Sb-Te5 | 93.2 |
|  |  |  | Te4-Ge-Te5 | 90.47 |  |  | Te4-Sb-Te5 | 83.8 |
|  | Ge-Te6 | 2.97 | Te1-Ge-Te6 | **89.1** | Sb-Te6 | 3.16 | Te1-Sb-Te6 | 93.2 |
|  |  |  | Te2-Ge-Te6 | **89.1** |  |  | Te2-Sb-Te6 | 93.2 |
|  |  |  | Te3-Ge-Te6 | 179.4 |  |  | Te3-Sb-Te6 | 175.9 |
|  |  |  | Te4-Ge-Te6 | 90.5 |  |  | Te4-Sb-Te6 | **83.8** |
|  |  |  | Te5-Ge-Te6 | 90.0 |  |  | Te5-Sb-Te6 | **83.8** |
|  | Ge-Te7 | 5.06 | Te1-Ge-Te7 | 124.3 | Sb-Te7 | 5.17 | Te1-Sb-Te7 | **54.7** |
|  |  |  | Te2-Ge-Te7 | 124.3 |  |  | Te2-Sb-Te7 | **54.7** |
|  |  |  | Te3-Ge-Te7 | 124.3 |  |  | Te3-Sb-Te7 | 124.9 |

|  | | Bond | θ | | | Bond | θ |
|---|---|---|---|---|---|---|---|
|  | | Te4-Ge-Te7 | **55.1** |  |  | Te4-Sb-Te7 | 125.2 |
|  | | Te5-Ge-Te7 | **55.1** |  |  | Te5-Sb-Te7 | 125.2 |
|  | | Te6-Ge-Te7 | **55.1** |  |  | Te6-Sb-Te7 | 59.2 |
|  | colspan Te1, CN=6, bonded to Ge ||| colspan Te2, CN=6, bonded to Ge and Sb ||||
|  | Bond | L | Bond | θ | Bond | L | Bond | θ |
|  | Te1-Ge1 | 2.97 | - | - | Te2-Ge1 | 2.95 | - | - |
|  | Te1-Ge2 | 2.97 | Ge1-Te1-Ge2 | 89.5 | Te2-Ge2 | 2.95 | Ge1-Te2-Ge2 | 91.3 |
|  | Te1-Ge3 | 2.97 | Ge1-Te1-Ge3 | 89.5 | Te2-Ge3 | 2.95 | Ge1-Te2-Ge3 | 91.3 |
|  |  |  | Ge2-Te1-Ge3 | 90.5 |  |  | Ge2-Te2-Ge3 | 91.3 |
|  | Te1-Ge4 | 2.97 | Ge1-Te1-Ge4 | 180.0 | Te2-Sb1 | 3.16 | Ge1-Te2-Sb1 | 174.8 |
|  |  |  | Ge2-Te1-Ge4 | 90.5 |  |  | Ge2-Te2-Sb1 | 92.3 |
|  |  |  | Ge3-Te1-Ge4 | 90.5 |  |  | Ge3-Te2-Sb1 | 92.3 |
|  | Te1-Ge5 | 2.97 | Ge1-Te1-Ge5 | 90.5 | Te2-Sb2 | 3.16 | Ge1-Te2-Sb2 | 92.3 |
|  |  |  | Ge2-Te1-Ge5 | 180.0 |  |  | Ge2-Te2-Sb2 | 174.8 |
|  |  |  | Ge3-Te1-Ge5 | 89.5 |  |  | Ge3-Te2-Sb2 | 92.3 |
|  |  |  | Ge4-Te1-Ge5 | 89.5 |  |  | Sb1-Te2-Sb2 | 83.8 |
|  | Te1-Ge6 | 2.97 | Ge1-Te1-Ge6 | 90.5 | Te2-Sb3 | 3.16 | Ge1-Te2-Sb3 | 92.3 |
|  |  |  | Ge2-Te1-Ge6 | **89.5** |  |  | Ge2-Te2-Sb3 | 92.3 |
|  |  |  | Ge3-Te1-Ge6 | 180.0 |  |  | Ge3-Te2-Sb3 | 174.8 |
|  |  |  | Ge4-Te1-Ge6 | **89.5** |  |  | Sb1-Te2-Sb3 | **83.8** |
|  |  |  | Ge5-Te1-Ge6 | 90.5 |  |  | Sb2-Te2-Sb3 | **83.8** |
|  | Te1-Ge7 | 5.15 | Ge1-Te1-Ge7 | 54.8 | Te2-Ge4 | 5.06 | Ge1-Te2-Ge4 | **55.7** |
|  |  |  | Ge2-Te1-Ge7 | **54.2** |  |  | Ge2-Te2-Ge4 | **55.7** |
|  |  |  | Ge3-Te1-Ge7 | 125.2 |  |  | Ge3-Te2-Ge4 | **55.7** |
|  |  |  | Ge4-Te1-Ge7 | 125.2 |  |  | Sb1-Te2-Ge4 | 129.5 |
|  |  |  | Ge5-Te1-Ge7 | 125.8 |  |  | Sb2-Te2-Ge4 | 129.5 |
|  |  |  | Ge6-Te1-Ge7 | 54.8 |  |  | Sb3-Te2-Ge4 | 129.5 |
|  | colspan Te3, CN=3, bonded to Sb ||| colspan - ||||
|  | Bond | L | Bond | θ | - | - | - | - |
|  | Te3-Sb1 | 2.99 | - | - | - | - | - | - |
|  | Te3-Sb2 | 2.99 | Sb1-Te3-Sb2 | 89.7 | - | - | - | - |
|  | Te3-Sb3 | 2.99 | Sb1-Te3-Sb3 | **89.7** | - | - | - | - |
|  |  |  | Sb2-Te3-Sb3 | **89.7** |  |  | - | - |
|  | Te3-Sb4 | 5.17 | Sb1-Te3-Sb4 | 124.9 | - | - | - | - |
|  |  |  | Sb2-Te3-Sb4 | **54.7** |  |  | - | - |
|  |  |  | Sb3-Te3-Sb4 | **54.7** |  |  | - | - |
| GST-326 | colspan Ge, CN=6 ||| colspan Sb, CN=6 ||||
|  | Bond | L | Bond | θ | Bond | L | Bond | θ |
|  | Ge-Te1 | 2.96 | - | - | Sb-Te1 | 2.99 | - | - |
|  | Ge-Te2 | 2.96 | Te1-Ge-Te2 | 90.6 | Sb-Te2 | 2.99 | Te1-Sb-Te2 | 89.6 |
|  | Ge-Te3 | 2.96 | Te1-Ge-Te3 | 89.4 | Sb-Te3 | 2.99 | Te1-Sb-Te3 | 89.6 |
|  |  |  | Te2-Ge-Te3 | 89.4 |  |  | Te2-Sb-Te3 | 89.6 |
|  | Ge-Te4 | 2.96 | Te1-Ge-Te4 | 90.6 | Sb-Te4 | 3.15 | Te1-Sb-Te4 | 175.9 |

| | | Te2-Ge-Te4 | 90.6 | | | Te2-Sb-Te4 | 93.3 |
|---|---|---|---|---|---|---|---|
| | | Te3-Ge-Te4 | 180 | | | Te3-Sb-Te4 | 93.3 |
| Ge-Te5 | 2.96 | Te1-Ge-Te5 | 89.4 | Sb-Te5 | 3.15 | Te1-Sb-Te5 | 93.3 |
| | | Te2-Ge-Te5 | 180 | | | Te2-Sb-Te5 | 175.9 |
| | | Te3-Ge-Te5 | 90.6 | | | Te3-Sb-Te5 | 93.3 |
| | | Te4-Ge-Te5 | 89.4 | | | Te4-Sb-Te5 | 83.6 |
| Ge-Te6 | 2.96 | Te1-Ge-Te6 | 180 | Sb-Te6 | 3.15 | Te1-Sb-Te6 | 93.3 |
| | | Te2-Ge-Te6 | **89.4** | | | Te2-Sb-Te6 | 93.3 |
| | | Te3-Ge-Te6 | 90.6 | | | Te3-Sb-Te6 | 175.9 |
| | | Te4-Ge-Te6 | **89.4** | | | Te4-Sb-Te6 | **83.6** |
| | | Te5-Ge-Te6 | 90.6 | | | Te5-Sb-Te6 | **83.6** |
| Ge-Te7 | 5.06 | Te1-Ge-Te7 | 124.8 | Sb-Te7 | 5.16 | Te1-Sb-Te7 | **54.6** |
| | | Te2-Ge-Te7 | 124.8 | | | Te2-Sb-Te7 | **54.6** |
| | | Te3-Ge-Te7 | **55.2** | | | Te3-Sb-Te7 | 124.8 |
| | | Te4-Ge-Te7 | 124.8 | | | Te4-Sb-Te7 | 125.3 |
| | | Te5-Ge-Te7 | **55.2** | | | Te5-Sb-Te7 | 125.3 |
| | | Te6-Ge-Te7 | **55.2** | | | Te6-Sb-Te7 | 59.3 |
| Te1, CN=6, bonded to Ge | | | | Te2, CN=6, bonded to Ge and Sb | | | |
| Bond | $L$ | Bond | $\theta$ | Bond | $L$ | Bond | $\theta$ |
| Te1-Ge1 | 2.96 | - | - | Te2-Ge1 | 2.94 | - | - |
| Te1-Ge2 | 2.96 | Ge1-Te1-Ge2 | 90.6 | Te2-Ge2 | 2.94 | Ge1-Te2-Ge2 | 91.2 |
| Te1-Ge3 | 2.96 | Ge1-Te1-Ge3 | 90.6 | Te2-Ge3 | 2.94 | Ge1-Te2-Ge3 | 91.2 |
| | | Ge2-Te1-Ge3 | 90.6 | | | Ge2-Te2-Ge3 | 91.2 |
| Te1-Ge4 | 2.97 | Ge1-Te1-Ge4 | 179.7 | Te2-Sb1 | 3.15 | Ge1-Te2-Sb1 | 92.5 |
| | | Ge2-Te1-Ge4 | 89.6 | | | Ge2-Te2-Sb1 | 92.5 |
| | | Ge3-Te1-Ge4 | 89.6 | | | Ge3-Te2-Sb1 | 174.8 |
| Te1-Ge5 | 2.97 | Ge1-Te1-Ge5 | 89.6 | Te2-Sb2 | 3.15 | Ge1-Te2-Sb2 | 92.5 |
| | | Ge2-Te1-Ge5 | 179.7 | | | Ge2-Te2-Sb2 | 174.8 |
| | | Ge3-Te1-Ge5 | 89.6 | | | Ge3-Te2-Sb2 | 92.5 |
| | | Ge4-Te1-Ge5 | 90.2 | | | Sb1-Te2-Sb2 | 83.6 |
| Te1-Ge6 | 2.97 | Ge1-Te1-Ge6 | **89.6** | Te2-Sb3 | 3.15 | Ge1-Te2-Sb3 | 174.8 |
| | | Ge2-Te1-Ge6 | **89.6** | | | Ge2-Te2-Sb3 | 92.5 |
| | | Ge3-Te1-Ge6 | 179.7 | | | Ge3-Te2-Sb3 | 92.5 |
| | | Ge4-Te1-Ge6 | 90.2 | | | Sb1-Te2-Sb3 | **83.6** |
| | | Ge5-Te1-Ge6 | 90.2 | | | Sb2-Te2-Sb3 | **83.6** |
| Te1-Ge7 | 5.15 | Ge1-Te1-Ge7 | **55.2** | Te2-Ge4 | 5.06 | Ge1-Te2-Ge4 | **55.6** |
| | | Ge2-Te1-Ge7 | **55.2** | | | Ge2-Te2-Ge4 | **55.6** |
| | | Ge3-Te1-Ge7 | **55.2** | | | Ge3-Te2-Ge4 | **55.6** |
| | | Ge4-Te1-Ge7 | 125.1 | | | Sb1-Te2-Ge4 | 129.6 |
| | | Ge5-Te1-Ge7 | 125.1 | | | Sb2-Te2-Ge4 | 129.6 |
| | | Ge6-Te1-Ge7 | 125.1 | | | Sb3-Te2-Ge4 | 129.6 |
| Te3, CN=3, bonded to Sb | | | | - | | | |
| Bond | $L$ | Bond | $\theta$ | - | - | - | - |

|  |  |  |  |  |  |  |  |  |
|---|---|---|---|---|---|---|---|---|
|  | Te3-Sb1 | 2.99 | - | - | - | - | - | - |
|  | Te3-Sb2 | 2.99 | Sb1-Te3-Sb2 | 89.6 | - | - | - | - |
|  | Te3-Sb3 | 2.99 | Sb1-Te3-Sb3 | **89.6** | - | - | - | - |
|  |  |  | Sb2-Te3-Sb3 | **89.6** |  |  | - | - |
|  | Te3-Sb4 | 5.18 | Sb1-Te3-Sb4 | **54.6** | - | - | - | - |
|  |  |  | Sb2-Te3-Sb4 | **54.6** |  |  | - | - |
|  |  |  | Sb3-Te3-Sb4 | 124.8 |  |  | - | - |
| Sb₂Te₃ | \multicolumn{4}{c}{Sb, CN=6} | \multicolumn{4}{c}{Te1, CN=6, bonded to Sb} |

| Material | Bond | L | Bond | θ | Bond | L | Bond | θ |
|---|---|---|---|---|---|---|---|---|
|  | Sb-Te1 | 3.03 | - | - | Te1-Sb1 | 3.20 | - | - |
|  | Sb-Te2 | 3.03 | Te1-Sb-Te2 | 91.4 | Te1-Sb2 | 3.20 | Sb1-Te1-Sb2 | 85.4 |
|  | Sb-Te3 | 3.03 | Te1-Sb-Te3 | 91.4 | Te1-Sb3 | 3.20 | Sb1-Te1-Sb3 | 94.6 |
|  |  |  | Te2-Sb-Te3 | 91.4 |  |  | Sb2-Te1-Sb3 | 94.6 |
|  | Sb-Te4 | 3.20 | Te1-Sb-Te4 | 175.8 | Te1-Sb4 | 3.20 | Sb1-Te1-Sb4 | 85.4 |
|  |  |  | Te2-Sb-Te4 | 91.5 |  |  | Sb2-Te1-Sb4 | 85.4 |
|  |  |  | Te3-Sb-Te4 | 91.5 |  |  | Sb3-Te1-Sb4 | 180 |
|  | Sb-Te5 | 3.20 | Te1-Sb-Te5 | 91.5 | Te1-Sb5 | 3.20 | Sb1-Te1-Sb5 | 94.6 |
|  |  |  | Te2-Sb-Te5 | 175.8 |  |  | Sb2-Te1-Sb5 | 180 |
|  |  |  | Te3-Sb-Te5 | 91.5 |  |  | Sb3-Te1-Sb5 | 85.4 |
|  |  |  | Te4-Sb-Te5 | 85.4 |  |  | Sb4-Te1-Sb5 | 94.6 |
|  | Sb-Te6 | 3.20 | Te1-Sb-Te6 | 91.5 | Te1-Sb6 | 3.20 | Sb1-Te1-Sb6 | 180 |
|  |  |  | Te2-Sb-Te6 | 91.5 |  |  | Sb2-Te1-Sb6 | 94.6 |
|  |  |  | Te3-Sb-Te6 | 175.8 |  |  | Sb3-Te1-Sb6 | **85.4** |
|  |  |  | Te4-Sb-Te6 | **85.4** |  |  | Sb4-Te1-Sb6 | 94.6 |
|  |  |  | Te5-Sb-Te6 | **85.4** |  |  | Sb5-Te1-Sb6 | **85.4** |
|  | Sb-Te7 | 5.29 | Te1-Sb-Te7 | **55.1** | Te1-Sb7 | 5.39 | Sb1-Te1-Sb7 | **53.6** |
|  |  |  | Te2-Sb-Te7 | **55.1** |  |  | Sb2-Te1-Sb7 | **53.6** |
|  |  |  | Te3-Sb-Te7 | 126.9 |  |  | Sb3-Te1-Sb7 | 60.1 |
|  |  |  | Te4-Sb-Te7 | 124.8 |  |  | Sb4-Te1-Sb7 | 119.9 |
|  |  |  | Te5-Sb-Te7 | 124.8 |  |  | Sb5-Te1-Sb7 | 126.4 |
|  |  |  | Te6-Sb-Te7 | 57.3 |  |  | Sb6-Te1-Sb7 | 126.4 |

|  | \multicolumn{4}{c}{Te3, CN=3, bonded to Sb} | \multicolumn{4}{c}{-} |
|---|---|---|---|---|---|---|---|---|
|  | Bond | L | Bond | θ | - | - | - | - |
|  | Te3-Sb1 | 3.03 | - | - | - | - | - | - |
|  | Te3-Sb2 | 3.03 | Sb1-Te3-Sb2 | 91.4 | - | - | - | - |
|  | Te3-Sb3 | 3.03 | Sb1-Te3-Sb3 | **91.4** | - | - | - | - |
|  |  |  | Sb2-Te3-Sb3 | **91.4** |  |  | - | - |
|  | Te3-Sb4 | 5.29 | Sb1-Te3-Sb4 | **55.1** | - | - | - | - |
|  |  |  | Sb2-Te3-Sb4 | **55.1** |  |  | - | - |
|  |  |  | Sb3-Te3-Sb4 | 126.9 |  |  | - | - |
| GeTe | \multicolumn{4}{c}{Ge, CN=6} | \multicolumn{4}{c}{Te1, CN=6, bonded to Ge} |

| Material | Bond | L | Bond | θ | Bond | L | Bond | θ |
|---|---|---|---|---|---|---|---|---|
|  | Ge-Te1 | 2.86 | - | - | Te1-Ge1 | 2.86 | - | - |

| | Ge-Te2 | 2.86 | Te1-Ge-Te2 | 95.4 | Te1-Ge2 | 2.86 | Ge1-Te1-Ge2 | 95.4 |
|---|---|---|---|---|---|---|---|---|
| | Ge-Te3 | 2.86 | Te1-Ge-Te3 | 95.4 | Te1-Ge3 | 2.86 | Ge1-Te1-Ge3 | 95.4 |
| | | | Te2-Ge-Te3 | 95.4 | | | Ge2-Te1-Ge3 | 95.4 |
| | Ge-Te4 | 3.24 | Te1-Ge-Te4 | 91.2 | Te1-Ge4 | 3.24 | Ge1-Te1-Ge4 | 170.2 |
| | | | Te2-Ge-Te4 | 91.2 | | | Ge2-Te1-Ge4 | 91.2 |
| | | | Te3-Ge-Te4 | 170.2 | | | Ge3-Te1-Ge4 | 91.2 |
| | Ge-Te5 | 3.24 | Te1-Ge-Te5 | 91.2 | Te1-Ge5 | 3.24 | Ge1-Te1-Ge5 | 91.2 |
| | | | Te2-Ge-Te5 | 170.2 | | | Ge2-Te1-Ge5 | 170.2 |
| | | | Te3-Ge-Te5 | 91.2 | | | Ge3-Te1-Ge5 | 91.2 |
| | | | Te4-Ge-Te5 | 81.4 | | | Ge4-Te1-Ge5 | 81.4 |
| | Ge-Te6 | 3.24 | Te1-Ge-Te6 | 170.2 | Te1-Ge6 | 3.24 | Ge1-Te1-Ge6 | 91.2 |
| | | | Te2-Ge-Te6 | 91.2 | | | Ge2-Te1-Ge6 | 91.2 |
| | | | Te3-Ge-Te6 | 91.2 | | | Ge3-Te1-Ge6 | 170.2 |
| | | | Te4-Ge-Te6 | **81.4** | | | Ge4-Te1-Ge6 | **81.4** |
| | | | Te5-Ge-Te6 | **81.4** | | | Ge5-Te1-Ge6 | **81.4** |
| | Ge-Te7 | 5.10 | Te1-Ge-Te7 | **58.7** | Te1-Ge7 | 5.10 | Ge1-Te1-Ge7 | **58.7** |
| | | | Te2-Ge-Te7 | **58.7** | | | Ge2-Te1-Ge7 | **58.7** |
| | | | Te3-Ge-Te7 | **58.7** | | | Ge3-Te1-Ge7 | **58.7** |
| | | | Te4-Ge-Te7 | 131.1 | | | Ge4-Te1-Ge7 | 131.1 |
| | | | Te5-Ge-Te7 | 131.1 | | | Ge5-Te1-Ge7 | 131.1 |
| | | | Te6-Ge-Te7 | 131.1 | | | Ge6-Te1-Ge7 | 131.1 |

**Table S6.** The corresponding bond angel $\theta_1$ (°) and $\theta_2$ (°) of six materials in **Figure 1**.

| GST-147 | | $\theta_1$ | $\theta_2$ | GST-124 | | $\theta_1$ | $\theta_2$ |
|---|---|---|---|---|---|---|---|
| | Ge, CN=6 | 88.7 | 53.2 | | Ge, CN=6 | 88.8 | 53.3 |
| | Sb, CN=6 | 84.7 | 54.7 | | Sb, CN=6 | 84.1 | 54.7 |
| | Te1, CN=6, bonded to Sb | 84.7 | 53.4 | | - | - | - |
| | Te2, CN=6, bonded to Ge and Sb | 84.3 | 55.0 | | Te2, CN=6, bonded to Ge and Sb | 84.1 | 55.0 |
| | Te3, CN=3, bonded to Sb | 90 | 54.8 | | Te3, CN=3, bonded to Sb | 89.9 | 54.7 |
| GST-225 | | $\theta_1$ | $\theta_2$ | GST-326 | | $\theta_1$ | $\theta_2$ |
| | Ge, CN=6 | 89.1 | 55.1 | | Ge, CN=6 | 89.4 | 55.2 |
| | Sb, CN=6 | 83.8 | 54.7 | | Sb, CN=6 | 83.6 | 54.6 |
| | Te1, CN=6, bonded to Ge | 89.5 | 54.2 | | Te1, CN=6, bonded to Ge | 89.6 | 55.2 |
| | Te2, CN=6, bonded to Ge and Sb | 83.8 | 55.7 | | Te2, CN=6, bonded to Ge and Sb | 83.6 | 55.6 |
| | Te3, CN=3, bonded to Sb | 89.7 | 54.7 | | Te3, CN=3, bonded to Sb | 89.6 | 54.6 |
| Sb$_2$Te$_3$ | | $\theta_1$ | $\theta_2$ | GeTe | | $\theta_1$ | $\theta_2$ |
| | - | - | - | | Ge, CN=6 | 81.4 | 58.7 |
| | Sb, CN=6 | 85.4 | 55.1 | | - | - | - |
| | Te1, CN=6, bonded to Sb | 85.4 | 53.6 | | Te1, CN=6, bonded to Ge | 81.4 | 58.7 |
| | - | - | - | | - | - | - |
| | Te3, CN=3, bonded to Sb | 91.4 | 55.1 | | - | - | - |

**Table S7**. Calculated bond lengths in $Sb_2Te_3$, GST-147, GST-124, GST-225, GST-326 and GeTe.

| | Bond length (Å) | | | |
|---|---|---|---|---|
| | Ge-Te (I) | Ge-Te (II) | Sb-Te (I) | Sb-Te (II) |
| $Sb_2Te_3$ | - | - | 3.03 | 3.20 |
| GST-147 | 2.96 | - | 2.99 | 3.15 |
| GST-124 | 2.96 | - | 2.99 | 3.15 |
| GST-225 | 2.96 | | 2.99 | 3.15 |
| GST-326 | 2.95 | - | 2.99 | 3.15 |
| GeTe | 2.86 | 3.24 | - | - |

**Table S8.** Parameters for the amorphous models. The unit of lattice constant is Å, while that of theoretical number density is $Å^{-3}$.

| Composition | Number of atoms in the cell | | | | Lattice constant | Atomic number density |
|---|---|---|---|---|---|---|
| | Ge | Sb | Te | Total | | |
| a-$Sb_2Te_3$ | - | 120 | 180 | 300 | 22.259 | 0.0272 |
| a-GST-147 | 23 | 92 | 161 | 276 | 21.658 | 0.0272 |
| a-GST-124 | 40 | 80 | 160 | 280 | 21.658 | 0.0276 |
| a-GST-225 | 60 | 60 | 150 | 270 | 21.270 | 0.0281 |
| a-GST-326 | 75 | 50 | 150 | 275 | 21.258 | 0.0286 |
| a-GeTe | 150 | 150 | - | 300 | 21.307 | 0.0310 |

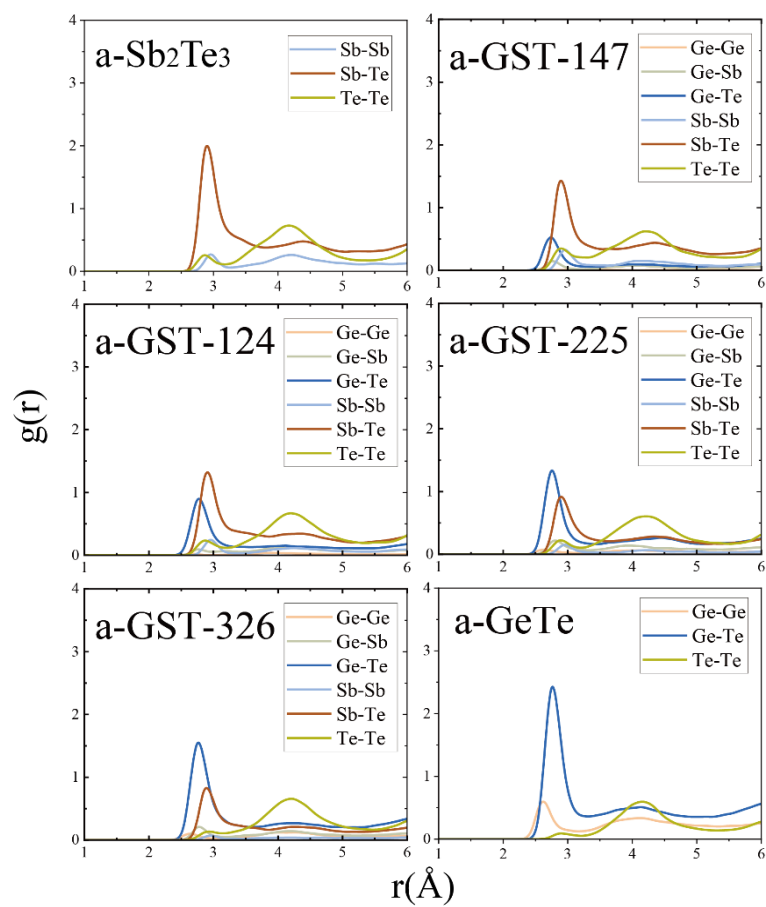

**Figure S9.** GGA-calculated partial pair distribution functions $g(r)$ for a-Sb$_2$Te$_3$, a-GST-147, a-GST-124, a-GST-225, a-GST-326 and a-GeTe at 300 K.

**Table S9.** The choices of bond length cutoff in counting the CNs. The first valley of $g(r)$ is always selected as the cutoff radius.

|       | GST-147 | GST-124 | GST-225 | GST-326 | Sb$_2$Te$_3$ | GeTe  |
|-------|---------|---------|---------|---------|--------------|-------|
| Ge-Ge | 3.156   | 2.95    | 3.249   | 3.049   | -            | 3.159 |
| Ge-Sb | 3.156   | 2.95    | 3.356   | 3.159   | -            | -     |
| Ge-Te | 3.391   | 3.35    | 3.446   | 3.547   | -            | 3.339 |
| Sb-Sb | 3.257   | 3.454   | 3.450   | 3.254   | 3.241        | -     |
| Sb-Te | 3.355   | 3.854   | 3.648   | 3.859   | 3.848        | -     |
| Te-Te | 3.350   | 3.154   | 3.249   | 3.159   | 3.167        | 3.24  |

## Supplementary Note 4. Radial distribution function (RDF($r$)), the average coordination number CN, pair-distribution function $g(r)$: Definitions and calculation formulae[2]

In amorphous materials, the most widely used method for calculating the coordination number of elements is through the integration of the radial distribution function (RDF($r$)), which is defined as follows.

$$RDF(r) = 4\pi\rho_0 g(r) r^2$$

where $\rho_0$ represents the atomic number density of the system, and $g(r)$ is the pair-distribution function

$$g(r) = \frac{N(r)}{4\pi r^2 dr}$$

Within the expression, $N(r)$ represents the total number of atomic pairs found in the shell $(r, r + dr)$; $4\pi r^2 dr$ is the volume of the spherical shell with radius $r$. Therefore, the formula for the average coordination number is as follows.

$$CN = 4\pi\rho_0 \int_0^{R_c} g(r) r^2 dr$$

**Supplementary Note 5. The definition of mobility gap[3]**

For amorphous materials, just as in crystals, electronic states may exist in bands separated by an energy gap. Unlike in crystals, however, the densities of states in the valence and conduction band differ and a joint density of states cannot be formulated. According to Mott and Davis[3], the bands in non-crystalline semiconductors can be divided into states localized near the band edge and delocalized extended states that are further away from the band edge. $E_c^m$ and $E_v^m$ indicate the energies at which this separation between localized and extended states occurs; their energetic distance is called the mobility gap $E_{gm}$.

**Supplementary Note 6. The calculation formula of the inverse participation ratio (IPR)**

The inverse participation ratio (IPR) is given by

$$\text{IPR} = \frac{\sum_i p_i^4}{\left(\sum_i p_i^2\right)^2}$$

where $p_i$ represents the projection components (projection values), which are the local density of states (LDOS) of electrons on different atoms. IPR reflects the localization of electron wave functions. Generally, a larger IPR value indicates a higher degree of electron localization, meaning that electrons are mainly concentrated on a few atoms; while a smaller IPR value suggests that the electrons are in a delocalized state, distributed over multiple atoms. Within the energy range of the conduction band and valence band, a higher IPR indicates that the electronic states at that energy are more localized, possibly related to impurity states or defect states. For localized electrons, their wave functions are mainly confined to a local region, while the wave functions of delocalized electrons spread over a broader region, resulting in a decrease in the IPR value.